\pdfoutput=1
\documentclass[11pt]{article}

\usepackage[final]{acl}

\usepackage{times}
\usepackage{latexsym}

\usepackage[T1]{fontenc}
\usepackage[utf8]{inputenc}

\usepackage{microtype}

\usepackage{inconsolata}

\usepackage{graphicx}

\usepackage{hyperref}
\usepackage{url}
\usepackage[utf8]{inputenc} % allow utf-8 input
\usepackage[T1]{fontenc}    % use 8-bit T1 fonts
\usepackage{hyperref}       % hyperlinks
\usepackage{url}            % simple URL typesetting
\usepackage{booktabs}       % professional-quality tables
\usepackage{amsfonts}       % blackboard math symbols
\usepackage{nicefrac}       % compact symbols for 1/2, etc.
\usepackage{microtype}      % microtypography
\usepackage{xcolor}         % colors
\usepackage{multirow}
\usepackage{amsmath}
\usepackage{amsthm}
\usepackage{array}
\usepackage{amssymb}% http://ctan.org/pkg/amssymb
\usepackage{pifont}% http://ctan.org/pkg/pifont
\usepackage{graphicx} 
\usepackage{wrapfig}
\usepackage{subcaption}
\usepackage{caption}
\usepackage{enumitem}
\usepackage[export]{adjustbox}
\usepackage{multibib}
\usepackage{ulem}
\usepackage{graphicx}
\usepackage{tabularx}

\title{FusionDTI: Fine-grained Binding Discovery with Token-level Fusion for Drug-Target Interaction}

\author{
Zhaohan Meng\textsuperscript{1},
Zaiqiao Meng\textsuperscript{1}\thanks{~Corresponding author},
Ke Yuan\textsuperscript{2,3} \&
Iadh Ounis\textsuperscript{1}\\
\textsuperscript{1}School of Computing Science \quad
\textsuperscript{2}School of Cancer Sciences \\
\textsuperscript{3}Cancer Research UK Scotland Institute \\
University of Glasgow, United Kingdom \\
\texttt{\{z.meng.3\}@research.gla.ac.uk}\\\vspace{1mm}
\texttt{\{zaiqiao.meng, ke.yuan, iadh.ounis\}@glasgow.ac.uk} \\
\parbox{0.03\textwidth}{\includegraphics[width=\linewidth]{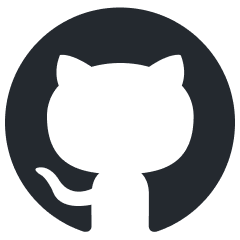}}\hspace{1mm}\href{https://github.com/ZhaohanM/FusionDTI}{\texttt{Source Code}}\hspace{10mm}
\parbox{0.03\textwidth}{\includegraphics[width=\linewidth]{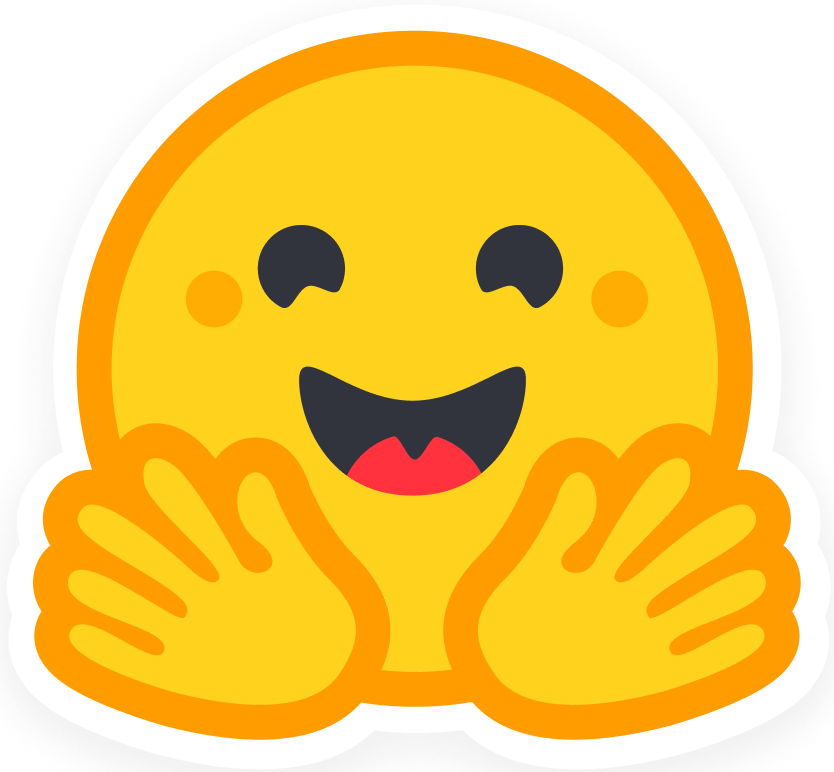}}\hspace{1mm}\href{https://huggingface.co/spaces/Zhaohan-Meng/FusionDTI}{\texttt{Demo Space}}\vspace{-3mm}
}
\begin{document}

\maketitle
\begin{abstract}
Predicting drug-target interaction (DTI) is critical in the drug discovery process. Despite remarkable advances in recent DTI models through the integration of representations from diverse drug and target encoders, such models often struggle to capture the fine-grained interactions between drugs and protein, i.e. the binding of specific drug atoms (or substructures) and key amino acids of proteins, which is crucial for understanding the binding mechanisms and optimising drug design. To address this issue, this paper introduces a novel model, called FusionDTI, which uses a token-level \textbf{Fusion} module to effectively learn fine-grained information for \textbf{D}rug-\textbf{T}arget \textbf{I}nteraction. In particular, our FusionDTI model uses the SELFIES representation of drugs to mitigate sequence fragment invalidation and incorporates the structure-aware (SA) vocabulary of target proteins to address the limitation of amino acid sequences in structural information, additionally leveraging pre-trained language models extensively trained on large-scale biomedical datasets as encoders to capture the complex information of drugs and targets. Experiments on three well-known benchmark datasets show that our proposed FusionDTI model achieves the best performance in DTI prediction compared with eight existing state-of-the-art baselines. Furthermore, our case study indicates that FusionDTI could highlight the potential binding sites, enhancing the explainability of the DTI prediction.

% \footnote{The complete code and datasets are available in the software section of the submission.}
\end{abstract}

\section{Introduction}

\begin{figure*}[hbpt!]
    \centering
    \includegraphics[width=0.85\linewidth]{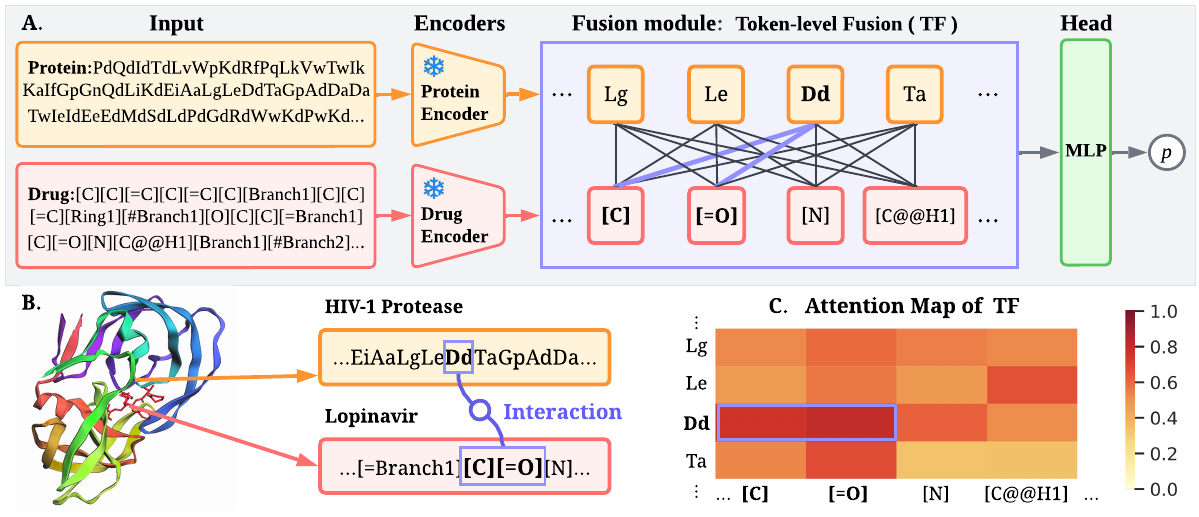}
    \caption{\textbf{A}. Illustration of the FusionDTI model: frozen encoder, fusion module and classifier. The token-level fusion (TF) focuses on fine-grained interactions between tokens within and across sequences. \textbf{B}. This is a token-level interaction instance of HIV-1 protease and lopinavir. Lopinavir forms a hydrogen bond with residue Dd (Asp25) in the active site of the protease via its ketone molecule ([C][=O]). \textbf{C}. The attention map of TF visualises the weight between tokens, indicating the contribution of each drug atom and residue to the final prediction result.}
    \label{fig:Proposed model}
\end{figure*}

The task of predicting drug-target interactions (DTI) plays a pivotal role in the drug discovery progress, as it helps identify potential therapeutic effects of drugs on biological targets facilitating the development of effective treatments~\citep{askr2023deep}. DTI fundamentally relies on the binding of specific drug atoms (or substructures) and key amino acids of proteins~\citep{schenone2013target}. In particular, each binding site is an interaction between a single amino acid and a single drug atom, which we refer to as a fine-grained interaction. For instance, Figure~\ref{fig:Proposed model} B demonstrates the interaction between \textit{HIV-1 protease} and the drug \textit{lopinavir}. A critical component of this interaction is the formation of a hydrogen bond between a ketone group in lopinavir (represented in the SELFIES~\citep{krenn2022selfies} notation as [C][=O]) and the side chain of an aspartate residue Asp25 (i.e. Dd) within the protease~\citep{brik2003hiv,chandwani2008lopinavir}. Therefore, capturing such fine-grained interaction information during the fusion of drug and target representations is crucial for building effective DTI prediction models~\citep{wu2022bridgedpi,peng2024bindti,zeng2024cat}.

To obtain representations of drugs and targets for the DTI task, some previous studies~\citep{lee2019deepconv,nguyen2021graphdta} have used graph neural networks (GNNs) or convolutional neural networks (CNNs) using a fixed-size window, potentially leading to a loss of contextual information, especially when drugs and targets are in a long-term sequence. These models directly concatenate the representations together to make predictions without considering fine-grained interactions. More recently, some computational models~\citep{huang2021moltrans,bai2023interpretable} employed the fusion module (e.g. Deep Interactive Inference Network (DIIN)~\citep{gong2017natural} and Bilinear Attention Network (BAN)~\citep{kim2018bilinear}) to obtain fine-grained interaction information and the 3-mer approach that binds three amino acids together as a target binding site to address the lack of structural information in the amino acid sequence. While useful for highlighting possible regions of interaction, these models do not offer the sufficient granularity needed to gauge the specifics of binding sites, as each binding site only contains one residue~\citep{schenone2013target}. Therefore, obtaining contextual representations of drugs and targets and capturing fine-grained interaction information for DTI remains challenging.

To address these challenges, we propose a novel model (called FusionDTI) with a Token-level Fusion (TF) module for an effective learning of fine-grained interactions between drugs and targets. In particular, our FusionDTI model utilises two pre-trained language models (PLMs), namely Saport~\citep{su2023saprot} as the protein encoder that is able to integrate both residue tokens with structure token; and SELFormer~\citep{yuksel2023selformer} as the drug encoder to ensure that each drug is valid and contains structural information. To effectively learn fine-grained information from these contextual representations of drugs and targets, we explore two strategies for the TF module, i.e.  Bilinear Attention Network (BAN)~\citep{kim2018bilinear} and Cross Attention Network (CAN)~\citep{li2021selfdoc,vaswani2017attention}, to find the best approach for integrating the rich contextual embeddings derived from Saport and SELFormer. We conduct a comprehensive performance comparison against eight existing state-of-the-art DTI prediction models. The results show that our proposed model achieves about 6\% accuracy improvement over the best baseline on the BindingDB dataset. The main contributions of our study are as follows:

\vspace{-6pt}
\begin{itemize}
    \item We propose FusionDTI, a novel model that leverages PLMs to encode drug SELFIES, as well as protein residues and structures for rich semantic representations and uses the token-level fusion to capture fine-grained interaction between drugs and targets effectively. 
\vspace{-3pt}
    \item We compare two TF modules: CAN and BAN and analyse the influence of fusion scales based on FusionDTI, demonstrating that CAN is superior for DTI prediction both in terms of effectiveness and efficiency.
\vspace{-3pt}    
    \item We conduct a case study of three drug-target pairs by FusionDTI to evaluate whether potential binding sites would be highlighted for the DTI prediction explainability.
\end{itemize}

\section{Related Work} 

\subsection{Drug and Protein Representation}

For drug molecules, most existing methods represent the input by the Simplified Molecular Input Line Entry System (SMILES)~\citep{weininger1988smiles,weininger1989smiles}. However, SMILES suffers from numerous problems in terms of validity and robustness, and some valuable information about the drug structure may be lost which may prevent the model from efficiently mining the knowledge hidden in the data~\citep{krenn2022selfies}. To address the limitations of SMILES, we apply SELFIES, a string-based representation that circumvents the issue of robustness and that always generates valid molecular graphs for each character. 

Regarding proteins, the conventional approach uses amino acid sequences as model inputs~\citep{huang2021moltrans,bai2023interpretable}, overlooking the crucial structural information of the protein. Inspired by the SA vocabulary of SaProt~\citep{su2023saprot}, the SaProt enhances inputs by amalgamating each residue of the amino acid sequence with a 3D geometric feature that is obtained by encoding protein structure information using Foldseek~\citep{van2024fast}. This innovative combination offers richer protein representations through the SA vocabulary, contributing to the discovery of fine-grained interactions. 

% Our proposed model employs SELFIES for drug encoding and uses SaProt encoding for proteins to generate semantic representations for both drugs and targets.

\subsection{Molecular and Protein Language Models}

Molecular language models trained on the large-scale molecular corpus capture the subtleties of chemical structures and their biological activities, setting new standards in the encoding of chemical compounds achieving meaningful representations~\cite{ying2021transformers,rong2020self}. For example, MoLFormer~\citep{ross2022large} focused on leveraging the self-attention mechanism to interpret the complex, non-linear interactions within molecules, while SELFormer~\citep{yuksel2023selformer} employed SELFIES, ensuring valid and interpretable chemical structures.

% ChemBERTa-2~\citep{ahmad2022chemberta} used RoBERTa-based architectures to capture intricate molecular patterns, significantly enhancing the precision of property prediction. Subsequently,

Protein language models have revolutionized the way we understand and represent protein sequences, learning intricate patterns and features that define the protein functionality and interactions. ProtBERT~\citep{protbert} and ESM~\citep{lin2023evolutionary} applied a transformer architecture to protein sequences, capturing the complex relationships between amino acids. Saport~\citep{su2023saprot,su2024saprothub} further enhanced this approach by integrating SA vocabularies to provide protein structure information.

% These models leverage the vast corpus of biological sequence data, learning intricate patterns and features that define the protein functionality and interactions.   
% Furthermore, SaprotHub~\citep{su2024saprothub} offers a platform that enables biologists to train, deploy, and share protein models efficiently. Importantly, our FusionDTI is flexible enough to use each of them as a protein encoder.

\begin{figure*}[htbp!]
    \centering
    \includegraphics[width=\linewidth]{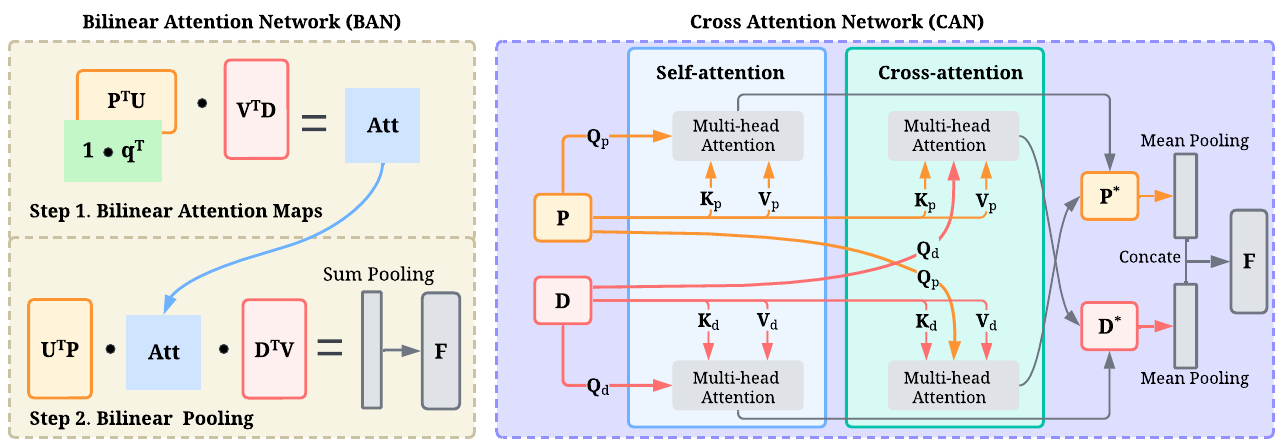}
    \caption{\textbf{BAN:} In step 1, the bilinear attention map is obtained by a bilinear interaction modelling via transformation matrices. In step 2, the joint representation \(\mathbf{F}\) is generated using the attention map by bilinear pooling via the shared transformation matrices \(\mathbf{U}\) and \(\mathbf{V}\). \textbf{CAN:} It fuses protein and drug representations through multi-head, self-attention and cross-attention. Then fused representations \(\mathbf{P}^{*}\) and \(\mathbf{D}^{*}\) are concatenated into \(\mathbf{F}\) after mean pooling.}
    \label{fig:FusionModule}
\end{figure*}

\section{Methodology}

\subsection{Model Architecture}

Given a sequence-based input drug-target pair, the DTI prediction task aims to predict an interaction probability score \(p \in [0, 1]\) between the given drug-target pair, which is typically achieved through learning a joint representation \(\mathbf{F}\) space from the given sequence-based inputs. To address the DTI task and effectively capture fine-grained interaction, we proposed a novel model, called FusionDTI, which is a bi-encoder model~\cite{liu2021trans} with a fusion module that fuses the representations of drugs and targets. The overall framework of FusionDTI is illustrated in Figure~\ref{fig:Proposed model} A. In general, FusionDTI takes sequence-based inputs of drugs and targets, which are encoded into token-level representation vectors by two frozen encoders. Then, a fusion module fuses the representations to capture fine-grained binding information for a final prediction through a prediction head.

\textbf{Input}: The initial inputs of drugs and targets are string-based representations. For protein \(\mathcal{P} \), the SA vocabulary~\citep{su2023saprot,van2024fast} is employed, where each residue is replaced by one of 441 SA vocabularies that bind an amino acid to a 3D geometric feature to address the lack of structural information in amino acid sequences. For drug \(\mathcal{D} \), as mentioned in the previous section, we use the SELFIES, which is a formal syntax that always generates valid molecular graphs~\citep{krenn2022selfies}. We provide the steps and code to obtain SA and SELFIES in Appendix~\ref{subsec:obtain_sequence}.
 
\textbf{Encoder}: The proposed model contains two frozen encoders: Saport~\citep{su2023saprot} and SELFormer~\citep{yuksel2023selformer}, which generate a drug representation \(\mathbf{D}\) and a protein representation \(\mathbf{P}\) separately. It is of note that FusionDTI is flexible enough to easily replace encoders with other PLMs or address SELFIES or SA representations that are unavailable. Furthermore, \(\mathbf{D}\) and \(\mathbf{P}\) are stored in memory for later-stage online training.

\textbf{Fusion module}: In developing FusionDTI, we have investigated two options for the fusion module: BAN and CAN to fuse representations, as indicated in Figure~\ref{fig:FusionModule}. The CAN is utilised to fuse each pair as \(\mathbf{D}^{*}\) and \(\mathbf{P}^{*}\), and then concatenate them into one \(\mathbf{F}\) for fine-grained binding information. For BAN, we need to obtain bilinear attention maps and generate \(\mathbf{F}\) through the bilinear pooling layer.

\textbf{Prediction head}: Finally, we obtain the probability score \(p\) of the DTI prediction by a multilayer perceptron (MLP) classifier trained with the binary cross-entropy loss, i.e. \(p=\operatorname{MLP}(\mathbf{F})\).

Since the encoders and the fusion module constitute the key components of our FusionDTI model, we will describe them in detail in the following.

\subsection{Drug and Protein Encoders}

Employing sequences with detailed biological functions and structures is a critical step in exploring the fine-grained binding of drugs and targets. For drugs, SMILES is the most commonly used input sequence but suffers from invalid sequence segments and potential loss of structural information~\citep{krenn2022selfies}. To address the limitations, we transform SMILES into SELFIES, a formal grammar that generates a valid molecular graph for each element~\citep{krenn2022selfies}. Besides, to address the lack of structural information in the amino acid sequences, we utilise the SA sequence of targets to combine each amino acid with an SA vocabulary by Foldseek~\citep{van2024fast}.

PLMs have shown promising achievements in the biomedical domain leveraging transformers since they pay attention to contextual information and are pre-trained on large-scale biomedical databases. Therefore, we utilise Saport~\citep{su2023saprot} as a protein encoder to encode protein input \(\mathcal{P} \) of both the SA sequence and amino acid sequence. Meanwhile, SELFormer~\citep{yuksel2023selformer} is used as our drug encoder to encode the drug SELFIES input \(\mathcal{D} \). Then these encoded protein representation $\mathbf{P}$ and drug representation $\mathbf{D}$ are further used as inputs for the later fusion module (Subsection~\ref{subsec:fusion}). These rich contextual representations ensure that we can explore the fine-grained binding information effectively. To further justify this, we also compare our encoders with other existing protein language models (such as ESM-2~\citep{lin2023evolutionary}) and molecular language models (such as MoLFormer~\citep{ross2022large} and ChemBERTa-2~\citep{ahmad2022chemberta}), and the results can be found in Appendix~\ref{subsec:PLMs}.

\subsection{Fusion Module}
\label{subsec:fusion}

In order to capture the fine-grained binding information between a drug and a target, our FusionDTI model applies a fusion module to learn token-level interactions between the token representations of drugs and targets encoded by their respective encoders. As shown in Figure~\ref{fig:FusionModule}, two fusion modules are investigated to fuse representations: the Bilinear Attention Network~\citep{kim2018bilinear} and the Cross Attention Network~\citep{vaswani2017attention}.

\subsubsection{Bilinear Attention Network (BAN)}

Motivated by DrugBAN~\citep{bai2023interpretable}, our model considers BAN~\citep{kim2018bilinear} as an option to learn pairwise fine-grained interactions between drug \(\mathbf{D} \in \mathbb{R}^{m \times \phi}\) and target \(\mathbf{P} \in \mathbb{R}^{n \times \rho}\), denoted as FusionDTI-BAN. For BAN as indicated in Figure~\ref{fig:FusionModule}, bilinear attention maps are obtained by a bilinear interaction modelling to capture pairwise weights in step 1, and then the bilinear pooling layer to extract a joint representation \(\mathbf{F}\). The equation of BAN is shown below:

\vspace{-7pt}
\begingroup
\small
\begin{equation}
\label{eq:pooling_ban}
        \begin{split}
        \mathbf{F} &= \operatorname{BAN}(\mathbf{P}, \mathbf{D} ; Att) \\
                   &= \mathrm{SumPool}(\sigma(\mathbf{P}^\top\mathbf{U}) \cdot Att \cdot \sigma(\mathbf{D}^\top\mathbf{V}), s),
        \end{split}
\end{equation}
\endgroup

where \(\mathbf{U} \in \mathbb{R}^{n \times K}\) and \(\mathbf{V} \in \mathbb{R}^{m \times K}\) are transformation matrices for representations. \(\mathrm{SumPool}\) is an operation that performs a one-dimensional and non-overlapped sum pooling operation with stride \(s\) and \(\sigma(\cdot) \) denotes a non-linear activation function with \(\mathrm{ReLU}(\cdot)\). \(Att \in \mathbb{R}^{\rho \times \phi}\) represents the bilinear attention maps using the Hadamard product and matrix-matrix multiplication and is defined as:

\begingroup
\small
\begin{equation}
\label{eq:att_ban}
Att =((\mathbf{1} \cdot \mathbf{q}^\top) \circ \sigma(\mathbf{P}^\top\mathbf{U})) \cdot \sigma(\mathbf{V}^\top\mathbf{D}),
\end{equation}
\endgroup

Here, \(\mathbf{1} \in \mathbb{R}^{\rho}\) is a fixed all-ones vector, \(\mathbf{q} \in \mathbb{R}^K\) is a learnable weight vector and \(\circ\) denotes the Hadamard product. In this way, pairwise interactions contribute sub-structural pairs to predictions. 

BAN captures the token-level interactions between the protein and drug representations without considering the relationships within each sequence itself, which may limit its ability to understand deeper contextual dependencies.

\subsubsection{Cross Attention Network (CAN)}
\vspace{7pt}
Inspired by ProST~\citep{xu2023protst}, we also consider CAN as our fusion module to learn fine-grained interaction information of drugs and targets. We denote our FusionDTI model that uses a CAN fusion module as FusionDTI-CAN. By processing \( \mathbf{D} \in \mathbb{R}^{m \times h} \) and \( \mathbf{P} \in \mathbb{R}^{n \times h} \) separately, the fused drug \( \mathbf{D}^{*}\in \mathbb{R}^{m \times h}\) and target \( \mathbf{P}^{*} \in \mathbb{R}^{n \times h}\) representations are obtained. To synthesise the fine-grained joint representation \( \mathbf{F}\), we employ a pooling aggregation strategy for both \( \mathbf{D}^{*}\) and \( \mathbf{P}^{*}\)independently and then concatenate them as shown in Figure~\ref{fig:FusionModule}. The process is described by the following equation:

\vspace{-12pt}
\begingroup
\small
\begin{equation}
\mathbf{F} = \mathrm{Concat}[\mathrm{MeanPool}(\mathbf{D}^{*}
), \mathrm{MeanPool}(\mathbf{P}^{*})],
\end{equation}
\endgroup
\vspace{-18pt}

where \(\mathrm{MeanPool}\) calculates the element-wise mean of all tokens across the sequence dimension, and \(\mathrm{Concat}\) denotes the concatenation of the resulting mean vectors. In this context, the multi-head, self-attention and cross-attention mechanisms are used to refine the representations of each residue and atom as below:

\vspace{-12pt}
\begingroup
\small
\begin{equation}
\label{eq:d_fused}
    \mathbf{D}^{*} = \frac{1}{2} \left[ \textit{MHA}(\mathbf{Q}_{d}, \mathbf{K}_{d}, \mathbf{V}_{d}) + \textit{MHA}(\mathbf{Q}_{p}, \mathbf{K}_{d}, \mathbf{V}_{d}) \right],
\end{equation}
\endgroup

\vspace{-18pt}
\begingroup
\small
\begin{equation}
\label{eq:p_fused}
    \mathbf{P}^{*} = \frac{1}{2} \left[ \textit{MHA}(\mathbf{Q}_{p}, \mathbf{K}_{p}, \mathbf{V}_{p}) + \textit{MHA}(\mathbf{Q}_{d}, \mathbf{K}_{p}, \mathbf{V}_{p}) \right],
\end{equation}
\endgroup

where \(\mathbf{Q}_{d}, \mathbf{K}_{d}, \mathbf{V}_{d} \in \mathbb{R}^{m \times h} \) and \(\mathbf{Q}_{p}, \mathbf{K}_{p}, \mathbf{V}_{p} \in \mathbb{R}^{n \times h} \) are the queries, keys and values for drug and target protein, respectively. And \textit{MHA} denotes the Multi-head Attention mechanism. To guide this process, two distinct sets of projection matrices guide the attention mechanism as follows:

\vspace{-12pt}
\begingroup
\small
\begin{equation}
\label{eq:qkv_drug}
\mathbf{Q}_d = \mathbf{D}\mathbf{W}_q^d, \quad \mathbf{K}_d = \mathbf{D}\mathbf{W}_k^d, \quad \mathbf{V}_d = \mathbf{D}\mathbf{W}_v^d,
\end{equation}
\endgroup
\vspace{-17pt}
\begingroup
\small
\begin{equation}
\label{eq:qkv_protein}
\mathbf{Q}_p = \mathbf{P}\mathbf{W}_q^p, \quad \mathbf{K}_p = \mathbf{P}\mathbf{W}_k^p, \quad \mathbf{V}_p = \mathbf{P}\mathbf{W}_v^p,
\end{equation}
\endgroup
Here, the projection matrices \( \mathbf{W}_q^d, \mathbf{W}_k^d, \mathbf{W}_v^d \in \mathbb{R}^{h \times h} \) and \( \mathbf{W}_q^p, \mathbf{W}_k^p, \mathbf{W}_v^p \in \mathbb{R}^{h \times h} \) are used to derive the queries, keys and values, respectively. 

In summary, our CAN module combines multi-head, self-attention and cross-attention mechanisms to capture dependencies within individual sequences and between different sequences for a more nuanced understanding of interactions. In the results of Sections~\ref{subsec:BAN_vs_CAN} and ~\ref{subsec:fusion_scales}, we analyse and compare these two fusion strategies and different fusion scales in detail.

\section{Experimental Setup and Results}

\begin{table*}[t]
\centering
\resizebox{\textwidth}{!}{
\huge
\begin{tabular}{c  ccc  cc  ccc}
\toprule
& \multicolumn{3}{c}{BindingDB} & \multicolumn{2}{c}{Human} & \multicolumn{3}{c}{BioSNAP} \\
\cmidrule(r){2-4}  \cmidrule(r){5-6}  \cmidrule(r){7-9} Method & 
AUROC & AUPRC & Accuracy  & AUROC & AUPRC & AUROC & AUPRC & Accuracy \\ \midrule
SVM       &  .939$\pm$.001 & .928$\pm$.002 & .825$\pm$.004 & .940$\pm$.006 & .920$\pm$.009 & .862$\pm$.007 & .864$\pm$.004  & .777$\pm$.011\\
RF        & .942$\pm$.011 & .921$\pm$.016 & .880$\pm$.012 & .952$\pm$.011 & .953$\pm$.010 & .860$\pm$.005 & .886$\pm$.005  & .804$\pm$.005\\
DeepConv-DTI & .945$\pm$.002 & .925$\pm$.005 & .882$\pm$.007 & .980$\pm$.002 & .981$\pm$.002 & .886$\pm$.006 & .890$\pm$.006  & .805$\pm$.009\\
GraphDTA  & .951$\pm$.002 & .934$\pm$.002 & .888$\pm$.005 & .981$\pm$.001 & .982$\pm$.002 & .887$\pm$.008 & .890$\pm$.007  & .800$\pm$.007\\
MolTrans & .952$\pm$.002 & .936$\pm$.001 & .887$\pm$.006 & .980$\pm$.002 & .978$\pm$.003 & .895$\pm$.004 & .897$\pm$.005  & .825$\pm$.010\\
DrugBAN & .960$\pm$.001 & .948$\pm$.002 & .904$\pm$.004 & .982$\pm$.002 & .980$\pm$.003 & .903$\pm$.005 & .902$\pm$.004  & .834$\pm$.008\\
SiamDTI & .961$\pm$.002 & .945$\pm$.002 & .890$\pm$.006 & .970$\pm$.002 & .969$\pm$.003 & .912$\pm$.005 & .910$\pm$.003  & .855$\pm$.004\\
BioT5 & .963$\pm$.001 & .952$\pm$.001 & .907$\pm$.003 & \underline{.989$\pm$.001} & \underline{.985$\pm$.002} & \underline{.937$\pm$.001} & \underline{.937$\pm$.004} & \underline{.874$\pm$.001}\\ \midrule
FusionDTI-BAN & \underline{.975$\pm$.002} & \underline{.976$\pm$.002} & \underline{.933$\pm$.003} & .984$\pm$.002 & .984$\pm$.003 & .923$\pm$.002 & .921$\pm$.002  & .856$\pm$.001\\
FusionDTI-CAN & \textbf{.989$\pm$.002} & \textbf{.990$\pm$.002} & \textbf{.961$\pm$.002} & \textbf{.991$\pm$.002} & \textbf{.989$\pm$.002}  & \textbf{.951$\pm$.002} & \textbf{.952$\pm$.002} & \textbf{.889$\pm$.002} \\
\bottomrule
\end{tabular}}
\caption{\label{tab:in_domain}In-domain performance comparison of FusionDTI and the baselines on the BindingDB, Human and BioSNAP datasets (\textbf{Best}, \underline{Second Best}).}
\end{table*}

\begin{table*}[t]
\centering
\resizebox{\textwidth}{!}{
\huge
\begin{tabular}{c  ccc  cc  ccc}
\toprule
& \multicolumn{3}{c}{BindingDB} & \multicolumn{2}{c}{Human} & \multicolumn{3}{c}{BioSNAP} \\
\cmidrule(r){2-4}  \cmidrule(r){5-6}  \cmidrule(r){7-9} Method & 
AUROC & AUPRC & Accuracy & AUROC & AUPRC & AUROC & AUPRC & Accuracy\\ \midrule

SVM       &  .490$\pm$.015 & .460$\pm$.001 & .531$\pm$.009 & .621$\pm$.036 & .637$\pm$.009 & .602$\pm$.005 & .528$\pm$.005  & .513$\pm$.011\\
RF        & .493$\pm$.021 & .468$\pm$.023 & .535$\pm$.012 & .642$\pm$.011 & .663$\pm$.050 & .590$\pm$.015 & .568$\pm$.018  & .499$\pm$.004\\
GraphDTA  & .536$\pm$.015 & .496$\pm$.029 & .472$\pm$.009 & .822$\pm$.009 & .759$\pm$.006 & .618$\pm$.005 & .618$\pm$.008  & .535$\pm$.024\\
DeepConv-DTI & .527$\pm$.038 & .499$\pm$.035 & .490$\pm$.027 & .761$\pm$.016 & .628$\pm$.022 & .645$\pm$.022 & .642$\pm$.032  & .558$\pm$.025\\
MolTrans & .554$\pm$.024 & .511$\pm$.025 & .470$\pm$.004 & .810$\pm$.021 & .745$\pm$.034 & .621$\pm$.015 & .608$\pm$.022  & .546$\pm$.032\\
DrugBAN & .604$\pm$.027 & .570$\pm$.047 & .509$\pm$.021 & .833$\pm$.020 & .760$\pm$.031 & .685$\pm$.044 & .713$\pm$.041 & .565$\pm$.056\\
SiamDTI & .627$\pm$.027 & .571$\pm$.024 & .563$\pm$.033 & \textbf{.863$\pm$.019} & \underline{.807$\pm$.040} & .718$\pm$.055 & .725$\pm$.054 & .623$\pm$.070\\
BioT5 & .651$\pm$.002 & .653$\pm$.003 & .621$\pm$.005 & \underline{.856$\pm$.003} & \textbf{.853$\pm$.003} & .720$\pm$.008 & .718$\pm$.004 & .715$\pm$.009\\\midrule
FusionDTI-BAN & \underline{.659$\pm$.002} & \underline{.663$\pm$.002} & \underline{.633$\pm$.003} & .784$\pm$.002 & .790$\pm$.003 & \underline{.723$\pm$.002} & \underline{.721$\pm$.002}  & \underline{.734$\pm$.001}\\
FusionDTI-CAN & \textbf{.681$\pm$.005} & \textbf{.680$\pm$.012} & \textbf{.652$\pm$.005} & .801$\pm$.037 & .803$\pm$.032 & \textbf{.748$\pm$.021} & \textbf{.766$\pm$.017} & \textbf{.756$\pm$.012} \\
\bottomrule
\end{tabular}}
\caption{\label{tab:cross_domain}Cross-domain performance comparison of FusionDTI and the baselines on the BindingDB, Human and BioSNAP datasets (\textbf{Best}, \underline{Second Best}).}
\end{table*}

\subsection{Datasets and Baselines}
  
Three public DTI datasets, namely BindingDB~\citep{gilson2016bindingdb}, BioSNAP~\citep{zitnik2018biosnap} and Human~\citep{liu2015improving,chen2020transformercpi}, are used for evaluation, where each dataset is split into training, validation, and test sets with a 7:1:2 ratio using two different splitting strategies: in-domain and cross-domain. For the in-domain split, the datasets are randomly divided. For the cross-domain setting, the datasets are split such that the drugs and targets in the test set do not overlap with those in the training set, making it a more challenging scenario where models must generalise to novel drug-target interactions. Since DTI is a binary classification task, we use AUROC~\citep{bai2023interpretable,huang2021moltrans} and AUPRC~\citep{nguyen2021graphdta} as the major metrics to evaluate models' performance. In Appendix~\ref{subsec:performance_comparison}, we report other evaluation metrics, including F1-score, Sensitivity, Specificity, and Matthews Correlation Coefficient (MCC) to provide a more comprehensive assessment.

We compare FusionDTI with eight baseline models in the DTI prediction task. These models include two traditional machine learning methods such as SVM~\citep{cortes1995support} and Random Forest (RF)~\citep{ho1995random}, as well as five deep learning methods including DeepConv-DTI~\citep{lee2019deepconv}, GraphDTA~\citep{nguyen2021graphdta}, MolTrans~\citep{huang2021moltrans}, DrugBAN~\citep{bai2023interpretable} and SiamDTI~\citep{zhang2024cross}. In addition, we also include the BioT5~\citep{pei-etal-2023-biot5} model, which is a biomedical pre-trained language model that could directly predict the DTI. 

Furthermore, results on three additional benchmark datasets (DAVIS~\citep{davis2011comprehensive}, KIBA~\citep{tang2014making}, and DUD-E~\citep{mysinger2012directory}) are reported, with comparisons to 8 task-specific baselines~\citep{nga2025lantern,li2025min}. Further details regarding the datasets, baseline models, and the methodology for generating drug SELFIES and protein SA sequences are provided in Appendix~\ref{subsec:obtain_sequence}.

% The latter five models employ the same two-stage process whereby the drug and target features are initially extracted by specialised encoders before being integrated for prediction.

\subsection{Evaluation of DTI Prediction}

We start by comparing our FusionDTI model (FusionDTI-CAN and FusionDTI-BAN) with eight existing state-of-the-art baselines for DTI prediction on three widely used datasets. Table~\ref{tab:in_domain} reports the in-domain comparative results. In general, our FusionDTI-CAN model performs the best on all metrics across all three datasets. A key highlight from these results is the exceptional performance of FusionDTI-CAN on the BindingDB dataset, where FusionDTI-CAN demonstrates superior metrics across the board: an AUROC of 0.989, an AUPRC of 0.990, and an accuracy of 96.1\%. Note that the main difference between the FusionDTI-CAN model and others is the fusion strategy. Furthermore, despite FusionDTI-BAN and DrugBAN both utilising the same BAN module, FusionDTI-BAN consistently outperforms DrugBAN on all datasets.

However, in-domain classification using random splits holds limited practical significance. Thus, we also evaluate the more challenging cross-domain DTI prediction, where the training data and the test data contain distinct drugs and targets. This setting precludes the use of known drug or target features when making predictions on the test data. As shown in Table~\ref{tab:cross_domain}, the performance of all models is diminished compared to the in-domain setting due to the reduced availability of information. Nevertheless, the FusionDTI-CAN model demonstrates outstanding performance in cross-domain DTI prediction on the BindingDB and BioSNAP datasets, highlighting its robustness in predicting novel drug-target interactions. For instance, on the BindingDB dataset, FusionDTI-CAN achieves the highest metrics with an AUROC of 0.675 and an AUPRC of 0.676. This underscores the effectiveness of the model's fusion strategy in diverse and challenging scenarios. Similarly, despite sharing the BAN module, FusionDTI-BAN continues to outperform DrugBAN, further confirming the effectiveness of the FusionDTI framework in addressing cross-domain prediction challenges.

These findings highlight not only the substantial improvements of FusionDTI over existing approaches but also its effectiveness in capturing fine-grained information on DTI. The key to this success lies in FusionDTI’s token-level fusion module, which enables the model to consider fine-grained interactions for each drug-target pair. This fine-grained interaction information aligns closely with biomedical pathways, where binding events often depend on the specific atoms or substructures involved in interactions with residues. Therefore, the model’s ability to capture such fine-grained interactions significantly enhances its predictive performance for DTI.

\subsection{Comparison of the BAN and CAN}
\label{subsec:BAN_vs_CAN}

\begin{figure}[t]
\centering
\includegraphics[width=\linewidth]{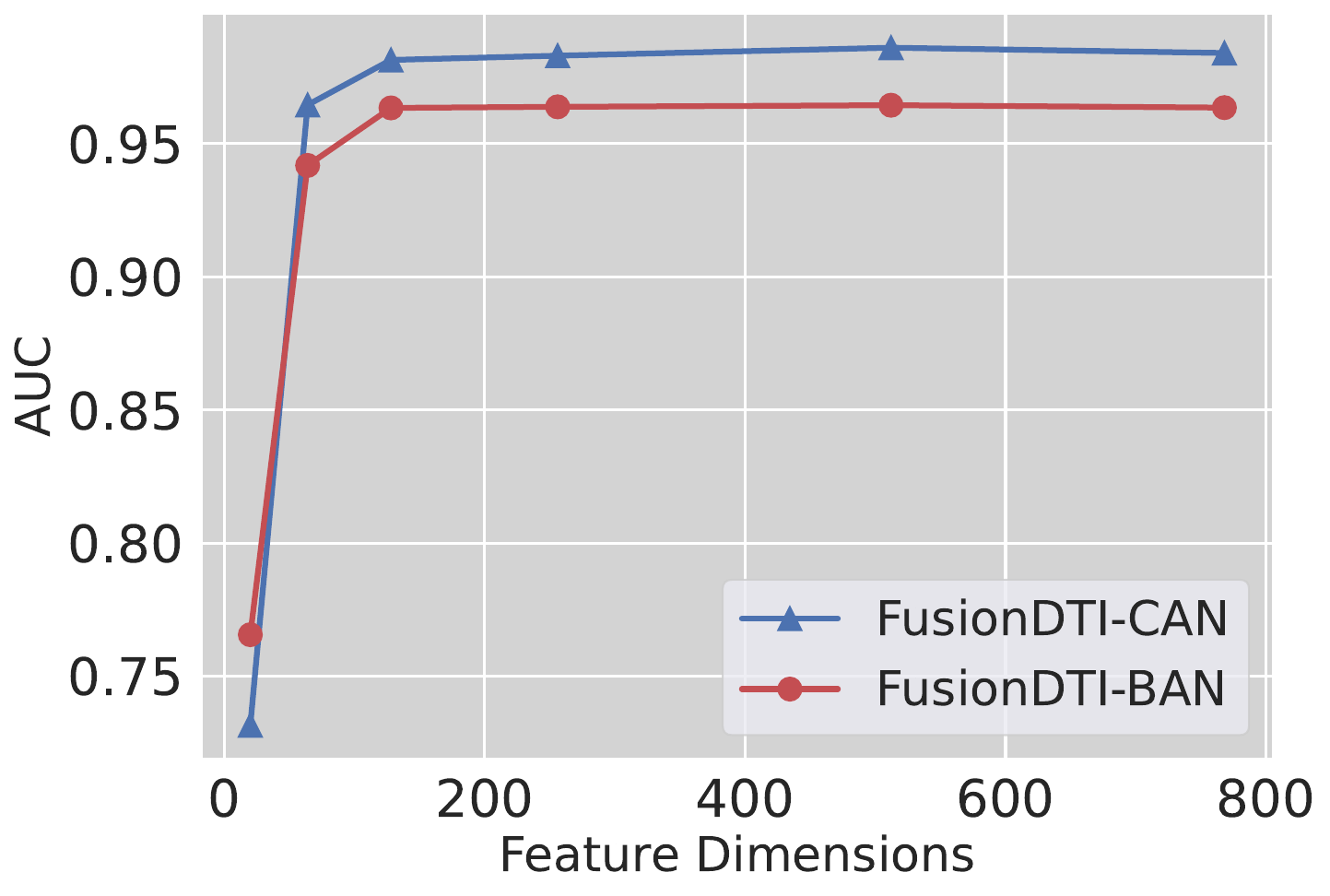}
\caption{\label{fig:Fusion_comparison}Performance comparison of two fusion strategies: BAN and CAN on the BindingDB.}
\end{figure}

There are two fusion strategies available: BAN and CAN, thus determining which one works better is a key step for establishing FusionDTI's prediction effectiveness. We perform a fair comparison involving the same encoders, classifier and dataset. As shown in Figure~\ref{fig:Fusion_comparison}, we compare BAN and CAN by employing two linear layers to adjust the feature dimensions of the drug and target representations. With the feature dimension increasing, the performance of FusionDTI-CAN continues to rise, while that of FusionDTI-BAN reaches a plateau. When the feature dimension is 512, both of the variants attain their peak positions with an AUC of 0.989 and 0.967, respectively. These results indicate that the CAN module seems to be better suited to the DTI prediction tasks and in capturing fine-grained interaction information. In contrast, BAN may not be able to fully capture fine-grained binding information between proteins and drugs, such as the specific interactions between the drug atoms and residues. Therefore, these findings suggest that the CAN strategy is more effective and adaptable to the complexities involved in DTI prediction, providing superior performance, especially as the feature dimension scales.

\subsection{Ablation Study}

\begin{figure}[t]
\centering
\scriptsize
\resizebox{0.97\columnwidth}{!}{%
\includegraphics[width=\linewidth]{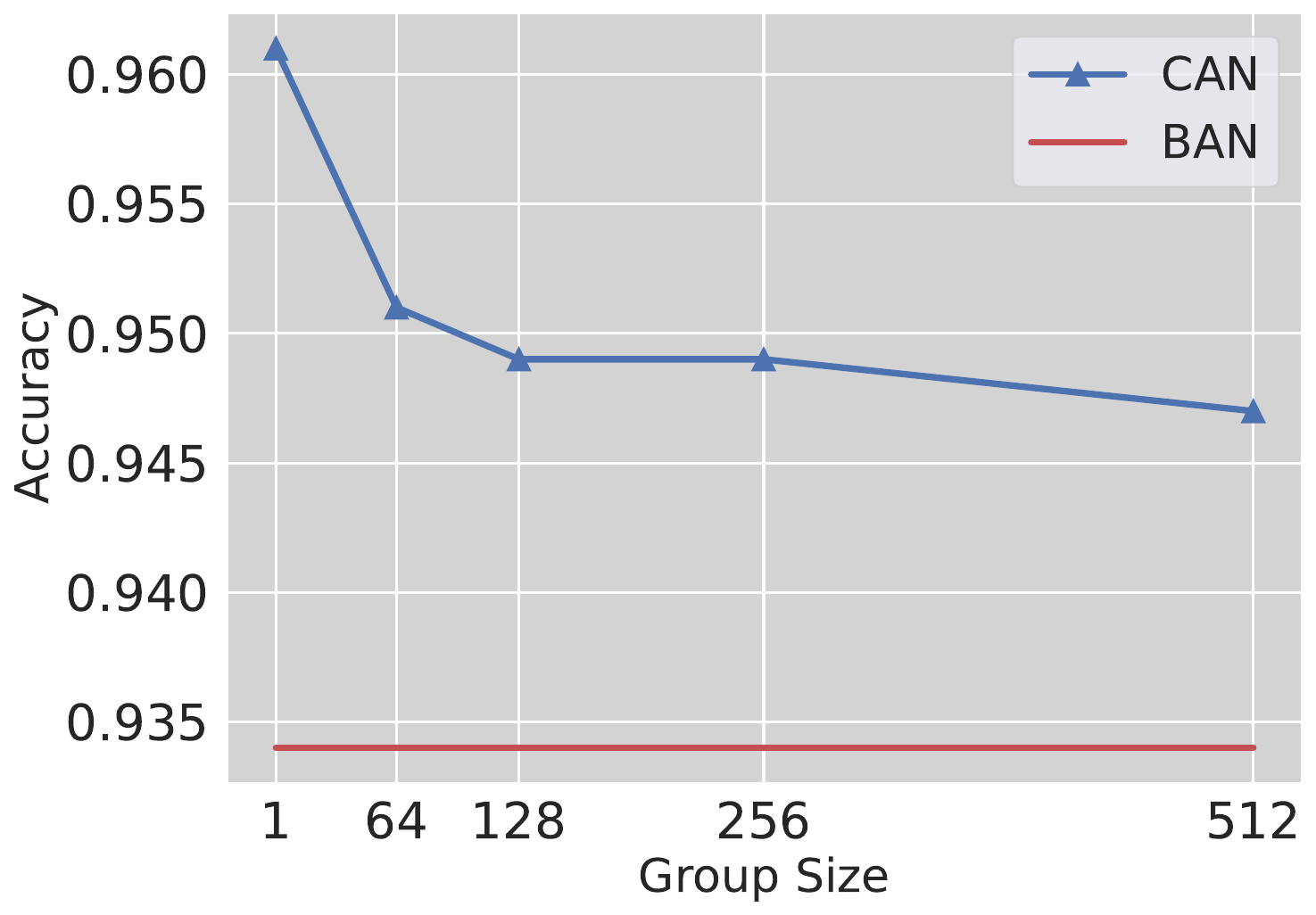}}
\caption{\label{fig:FusionDTI_Scales}Performance evaluation of fusion scales on the BindingDB dataset.}
\end{figure}

\begin{table}[t]
        \centering
        \footnotesize
        \begin{tabularx}{\columnwidth}{XXXX}
            \toprule
            CAN & AUC  & AUPRC & Accuracy\\ 
            \midrule
             \(\times\)& 0.954 & 0.963 & 0.894 \\ 
            \(\checkmark\) & 0.989 & 0.990 & 0.961 \\
            \bottomrule
        \end{tabularx}
        \caption{\label{tab:w/o_can}Ablation study of the CAN module on the BindingDB dataset.}
\end{table}

The fine-grained interaction of drug and target representations is critical in DTI as it directly impacts the model's ability to infer potential binding sites. For FusionDTI, this interaction is facilitated by the CAN module, which markedly enhances the predictive accuracy by capturing the fine-grained interaction information between the drugs and targets. Table~\ref{tab:w/o_can} demonstrates the impact of the CAN module on the prediction performance. When the fusion module is omitted, the model achieves an AUC of 0.954 and an accuracy of 0.894. Conversely, using the CAN module, there is a significant improvement, with the AUC increasing to 0.989 and the accuracy reaching 0.961. This highlights the effectiveness of the CAN module in improving the inference ability of FusionDTI.  In Appendix~\ref{subsec:Efficiency_Analysis} and~\ref{subsec:time_complexity}, we further compare time-consuming and time complexity with baselines.

\subsection{Analysis of Fusion Scales}
\label{subsec:fusion_scales}

In assessing fusion representations, it is critical to determine whether more fine-grained modelling enhances the predictive performance. Thus, we define a grouping function with the parameter \textbf{g} (Group size) for averaging tokens within each group before the CAN fusion module. The parameter \textbf{g}, representing the number of tokens per group, controls the granularity of the attention mechanism. Specifically, when \textbf{g} is set to 1, the fusion operates at the token level, where each token is considered independently. In contrast, when \textbf{g} is set to 512, the fusion occurs at a global level, considering the entire embedding as a single unit. We have the flexibility to control the fusion scale for the drug and protein representations, but the token length must be divisible by the group size. As shown in Figure~\ref{fig:FusionDTI_Scales}, as the number of tokens per group increases from 1 to 512 (Maximum Token Length), the performance of the FusionDTI model declines accordingly. This also aligns with the biomedical rules governing drug-protein interactions, where the principal factor influencing the binding is the interplay between the key atoms or substructures in the drug and primary residues in the protein. Furthermore, the CAN module outperforms BAN consistently at various scale settings, indicating that CAN better accesses the information between the drug and target. Consequently, this supports that the more detailed the interaction information obtained between the drugs and targets by the fusion module, the more beneficial it is for the enhancement of the model's prediction performance. 

\subsection{Case Study}

\begin{table}[t]
    \centering
    \small
    \begin{tabularx}{\columnwidth}{>{\raggedright\arraybackslash}X}
        \toprule
        \multicolumn{1}{c}{\textbf{Drug-Target Interactions}} \\
        \midrule
        \textbf{EZL - 6QL2:} \\
        \textbf{1}. sulfonamide oxygen - Leu198, Thr199 and Trp209; \\
        \textbf{2}. amino group - His94, His96, \textbf{His119} and Thr199; \\
        \textbf{3}. benzothiazole ring - Leu198, Thr200, \textbf{Tyr131}, Pro201 and \textbf{Gln92}; \\
        \textbf{4}. ethoxy group - \textbf{Gln135}; \\
        \midrule
        \textbf{9YA - 5W8L:} \\
        \textbf{1}. amino group of sulfonamide - Asp140, Glu191; \\
        \textbf{2}. sulfonamide oxygen - Asp140, Ile141 and \textbf{Val139}; \\
        \textbf{3}. carboxylic acid oxygens - Arg168, His192, \textbf{Asp194} and Thr247; \\
        \textbf{4}. biphenyl rings - Arg105, Asn137 and \textbf{Pro138}; \\
        \textbf{5}. hydrophobic contact - \textbf{Ala237}, \textbf{Tyr238} and \textbf{Leu322}; \\
        \midrule
        \textbf{EJ4 - 4N6H:} \\
        \textbf{1}. basic nitrogen of ligand - Asp128; \\
        \textbf{2}. hydrophobic pocket - Tyr308, Ile304 and \textbf{Tyr129}; \\
        \textbf{3}. water molecules - \textbf{Tyr129}, Met132, \textbf{Trp274}, Tyr308 and Lys214; \\
        \bottomrule
    \end{tabularx}
    \caption{FusionDTI predictions: \textbf{Bold} represents new predictions versus DrugBAN.}
    \label{tab:case}
\end{table}

\begin{figure}[t]
    \centering
    \scriptsize
    \resizebox{0.42\columnwidth}{!}{%
        \includegraphics{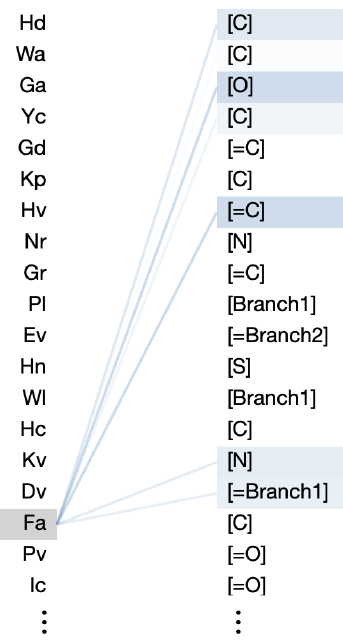}}
    \caption{EZL - 6QL2: Fine-grained interactions via attention visualization.}
    \label{fig:case_sample}
\end{figure}

A further strength of FusionDTI to enable explainability, which is critical for drug design efforts, is the visualisation of each token's contribution to the final prediction through cross-attention maps. To compare with the DrugBAN model, we examine three identical pairs of DTI from the Protein Data Bank (PDB)~\citep{berman2007worldwide}: (EZL - 6QL2~\citep{kazokaite2019engineered}, 9YA - 5W8L~\citep{rai2017discovery} and EJ4 - 4N6H~\citep{fenalti2014molecular}), which are excluded from the training data. As shown in Table~\ref{tab:case}, our proposed model predicts more binding sites existing in the PDB~\citep{berman2007worldwide} (in bold) by ranking the binding sites shown in the attention map. For instance, to predict the interaction of the drug EZL with the target 6QL2, our proposed model using BertViz~\citep{vig-2019-multiscale} highlights potential binding sites as illustrated in Figure~\ref{fig:case_sample}. Specifically, our CAN module is effective in capturing fine-grained binding information at the token level, as we have successfully predicted the novel binding between Gln92 and the benzothiazole ring~\citep{di2008carbonic}. In particular, we address the lack of structural information on protein sequences by employing the SA vocabulary, which matches each residue to a corresponding 3D feature via Foldseek~\citep{van2024fast}. This study highlights the effectiveness of FusionDTI in enhancing performance on the DTI task, thereby supporting more targeted and efficient drug development efforts. In Appendix~\ref{subsec:case_study}, we further investigate ten DTI pairs in non-small cell lung cancer (NSCLC) from PDB~\citep{waliany2025evolution}, highlighting predicted binding residues.

% In Appendix~\ref{subsec:case_study}, we present three pairs of prediction visualisations.

\section{Conclusions}

With the rapid increase of new diseases and the urgent need for innovative drugs, it is critical to capture fine-grained interactions, since the binding of specific drug atoms to the main amino acids is key to the DTI task. Despite some achievements, fine-grained interaction information is not effectively captured. To address this challenge, we introduce FusionDTI uses token-level fusion to effectively obtain fine-grained interaction information. Through experiments on three well-known datasets, we demonstrate that our proposed FusionDTI model outperforms eight state-of-the-art baselines, particularly in the more realistic cross-domain scenario. Additionally, we show that the attention weights of the token-level fusion module can highlight potential binding sites, providing a certain level of explainability.

% For future studies, we aim to investigate token-level interaction in more detail.

\clearpage
\section*{Limitations}

Even if our proposed model identifies potentially useful DTI, these predictions need to be validated by wet experiments, a time-consuming and expensive process. We have shown that FusionDTI is effective and efficient in screening for possible DTI in large-scale data as well as in locating potential binding sites in the process of drug design. However, it is not directly applicable to human medical therapy and other biomedical interactions because it lacks clinical validation and regulatory approval for medical use.

\bibliography{acl_latex}

\begin{thebibliography}{60}
\providecommand{\natexlab}[1]{#1}

\bibitem[{Ahmad et~al.(2022)Ahmad, Simon, Chithrananda, Grand, and
  Ramsundar}]{ahmad2022chemberta}
Walid Ahmad, Elana Simon, Seyone Chithrananda, Gabriel Grand, and Bharath
  Ramsundar. 2022.
\newblock Chemberta-2: Towards chemical foundation models.
\newblock \emph{arXiv preprint arXiv:2209.01712}.

\bibitem[{Askr et~al.(2023)Askr, Elgeldawi, Aboul~Ella, Elshaier, Gomaa, and
  Hassanien}]{askr2023deep}
Heba Askr, Enas Elgeldawi, Heba Aboul~Ella, Yaseen~AMM Elshaier, Mamdouh~M
  Gomaa, and Aboul~Ella Hassanien. 2023.
\newblock Deep learning in drug discovery: an integrative review and future
  challenges.
\newblock \emph{Artificial Intelligence Review}, 56(7):5975--6037.

\bibitem[{Bai et~al.(2023)Bai, Miljkovi{\'c}, John, and
  Lu}]{bai2023interpretable}
Peizhen Bai, Filip Miljkovi{\'c}, Bino John, and Haiping Lu. 2023.
\newblock Interpretable bilinear attention network with domain adaptation
  improves drug--target prediction.
\newblock \emph{Nature Machine Intelligence}, 5(2):126--136.

\bibitem[{Berman et~al.(2007)Berman, Henrick, Nakamura, and
  Markley}]{berman2007worldwide}
Helen Berman, Kim Henrick, Haruki Nakamura, and John~L Markley. 2007.
\newblock The worldwide protein data bank (wwpdb): ensuring a single, uniform
  archive of pdb data.
\newblock \emph{Nucleic acids research}, 35(suppl\_1):D301--D303.

\bibitem[{Brik and Wong(2003)}]{brik2003hiv}
Ashraf Brik and Chi-Huey Wong. 2003.
\newblock Hiv-1 protease: mechanism and drug discovery.
\newblock \emph{Organic \& biomolecular chemistry}, 1(1):5--14.

\bibitem[{Cao et~al.(2013)Cao, Xu, and Liang}]{cao2013propy}
Dong-Sheng Cao, Qing-Song Xu, and Yi-Zeng Liang. 2013.
\newblock propy: a tool to generate various modes of chou’s pseaac.
\newblock \emph{Bioinformatics}, 29(7):960--962.

\bibitem[{Chandwani and Shuter(2008)}]{chandwani2008lopinavir}
Ashish Chandwani and Jonathan Shuter. 2008.
\newblock Lopinavir/ritonavir in the treatment of hiv-1 infection: a review.
\newblock \emph{Therapeutics and clinical risk management}, 4(5):1023--1033.

\bibitem[{Chen et~al.(2020)Chen, Tan, Wang, Zhong, Liu, Yang, Luo, Chen, Jiang,
  and Zheng}]{chen2020transformercpi}
Lifan Chen, Xiaoqin Tan, Dingyan Wang, Feisheng Zhong, Xiaohong Liu, Tianbiao
  Yang, Xiaomin Luo, Kaixian Chen, Hualiang Jiang, and Mingyue Zheng. 2020.
\newblock Transformercpi: improving compound--protein interaction prediction by
  sequence-based deep learning with self-attention mechanism and label reversal
  experiments.
\newblock \emph{Bioinformatics}, 36(16):4406--4414.

\bibitem[{Cortes and Vapnik(1995)}]{cortes1995support}
Corinna Cortes and Vladimir Vapnik. 1995.
\newblock Support-vector networks.
\newblock \emph{Machine learning}, 20:273--297.

\bibitem[{Davis et~al.(2011)Davis, Hunt, Herrgard, Ciceri, Wodicka, Pallares,
  Hocker, Treiber, and Zarrinkar}]{davis2011comprehensive}
Mindy~I Davis, Jeremy~P Hunt, Sanna Herrgard, Pietro Ciceri, Lisa~M Wodicka,
  Gabriel Pallares, Michael Hocker, Daniel~K Treiber, and Patrick~P Zarrinkar.
  2011.
\newblock Comprehensive analysis of kinase inhibitor selectivity.
\newblock \emph{Nature biotechnology}, 29(11):1046--1051.

\bibitem[{Di~Fiore et~al.(2008)Di~Fiore, Pedone, Antel, Waldeck, Witte, Wurl,
  Scozzafava, Supuran, and De~Simone}]{di2008carbonic}
Anna Di~Fiore, Carlo Pedone, Jochen Antel, Harald Waldeck, Andreas Witte,
  Michael Wurl, Andrea Scozzafava, Claudiu~T Supuran, and Giuseppina De~Simone.
  2008.
\newblock Carbonic anhydrase inhibitors: the x-ray crystal structure of
  ethoxzolamide complexed to human isoform ii reveals the importance of thr200
  and gln92 for obtaining tight-binding inhibitors.
\newblock \emph{Bioorganic \& medicinal chemistry letters}, 18(8):2669--2674.

\bibitem[{Elnaggar et~al.(2021)Elnaggar, Heinzinger, Dallago, Rehawi, Yu,
  Jones, Gibbs, Feher, Angerer, Steinegger, Bhowmik, and Rost}]{protbert}
Ahmed Elnaggar, Michael Heinzinger, Christian Dallago, Ghalia Rehawi, Wang Yu,
  Llion Jones, Tom Gibbs, Tamas Feher, Christoph Angerer, Martin Steinegger,
  Debsindhu Bhowmik, and Burkhard Rost. 2021.
\newblock Prottrans: Towards cracking the language of lifes code through
  self-supervised deep learning and high performance computing.
\newblock \emph{IEEE Transactions on Pattern Analysis and Machine
  Intelligence}, pages 1--1.

\bibitem[{Fenalti et~al.(2014)Fenalti, Giguere, Katritch, Huang, Thompson,
  Cherezov, Roth, and Stevens}]{fenalti2014molecular}
Gustavo Fenalti, Patrick~M Giguere, Vsevolod Katritch, Xi-Ping Huang, Aaron~A
  Thompson, Vadim Cherezov, Bryan~L Roth, and Raymond~C Stevens. 2014.
\newblock Molecular control of $\delta$-opioid receptor signalling.
\newblock \emph{Nature}, 506(7487):191--196.

\bibitem[{Gao et~al.(2023)Gao, Qiang, Tan, Jia, Ren, Lu, Liu, Ma, and
  Lan}]{gao2023drugclip}
Bowen Gao, Bo~Qiang, Haichuan Tan, Yinjun Jia, Minsi Ren, Minsi Lu, Jingjing
  Liu, Wei-Ying Ma, and Yanyan Lan. 2023.
\newblock Drugclip: Contrastive protein-molecule representation learning for
  virtual screening.
\newblock \emph{Advances in Neural Information Processing Systems},
  36:44595--44614.

\bibitem[{Gilson et~al.(2016)Gilson, Liu, Baitaluk, Nicola, Hwang, and
  Chong}]{gilson2016bindingdb}
Michael~K Gilson, Tiqing Liu, Michael Baitaluk, George Nicola, Linda Hwang, and
  Jenny Chong. 2016.
\newblock Bindingdb in 2015: a public database for medicinal chemistry,
  computational chemistry and systems pharmacology.
\newblock \emph{Nucleic acids research}, 44(D1):D1045--D1053.

\bibitem[{Gong et~al.(2018)Gong, Luo, and Zhang}]{gong2017natural}
Yichen Gong, Heng Luo, and Jian Zhang. 2018.
\newblock Natural language inference over interaction space.
\newblock \emph{International Conference on Learning Representations}.

\bibitem[{Herrera-Ju{\'a}rez et~al.(2023)Herrera-Ju{\'a}rez, Serrano-G{\'o}mez,
  Bote-de Cabo, and Paz-Ares}]{herrera2023targeted}
Mercedes Herrera-Ju{\'a}rez, Cristina Serrano-G{\'o}mez, Helena Bote-de Cabo,
  and Luis Paz-Ares. 2023.
\newblock Targeted therapy for lung cancer: Beyond egfr and alk.
\newblock \emph{Cancer}, 129(12):1803--1820.

\bibitem[{Ho(1995)}]{ho1995random}
Tin~Kam Ho. 1995.
\newblock Random decision forests.
\newblock In \emph{Proceedings of 3rd international conference on document
  analysis and recognition}, volume~1, pages 278--282. IEEE.

\bibitem[{Huang et~al.(2021)Huang, Xiao, Glass, and Sun}]{huang2021moltrans}
Kexin Huang, Cao Xiao, Lucas~M Glass, and Jimeng Sun. 2021.
\newblock Moltrans: molecular interaction transformer for drug--target
  interaction prediction.
\newblock \emph{Bioinformatics}, 37(6):830--836.

\bibitem[{Kazokait{\.e} et~al.(2019)Kazokait{\.e}, Kairys, Smirnovien{\.e},
  Smirnov, Manakova, Tolvanen, Parkkila, and Matulis}]{kazokaite2019engineered}
Justina Kazokait{\.e}, Visvaldas Kairys, Joana Smirnovien{\.e}, Alexey Smirnov,
  Elena Manakova, Martti Tolvanen, Seppo Parkkila, and Daumantas Matulis. 2019.
\newblock Engineered carbonic anhydrase vi-mimic enzyme switched the structure
  and affinities of inhibitors.
\newblock \emph{Scientific reports}, 9(1):12710.

\bibitem[{Kim et~al.(2018)Kim, Jun, and Zhang}]{kim2018bilinear}
Jin-Hwa Kim, Jaehyun Jun, and Byoung-Tak Zhang. 2018.
\newblock Bilinear attention networks.
\newblock \emph{Advances in neural information processing systems}, 31.

\bibitem[{Krenn et~al.(2022)Krenn, Ai, Barthel, Carson, Frei, Frey, Friederich,
  Gaudin, Gayle, Jablonka et~al.}]{krenn2022selfies}
Mario Krenn, Qianxiang Ai, Senja Barthel, Nessa Carson, Angelo Frei, Nathan~C
  Frey, Pascal Friederich, Th{\'e}ophile Gaudin, Alberto~Alexander Gayle,
  Kevin~Maik Jablonka, and 1 others. 2022.
\newblock Selfies and the future of molecular string representations.
\newblock \emph{Patterns}, 3(10).

\bibitem[{Lee et~al.(2019)Lee, Keum, and Nam}]{lee2019deepconv}
Ingoo Lee, Jongsoo Keum, and Hojung Nam. 2019.
\newblock Deepconv-dti: Prediction of drug-target interactions via deep
  learning with convolution on protein sequences.
\newblock \emph{PLoS computational biology}, 15(6):e1007129.

\bibitem[{Li et~al.(2021)Li, Gu, Kuen, Morariu, Zhao, Jain, Manjunatha, and
  Liu}]{li2021selfdoc}
Peizhao Li, Jiuxiang Gu, Jason Kuen, Vlad~I Morariu, Handong Zhao, Rajiv Jain,
  Varun Manjunatha, and Hongfu Liu. 2021.
\newblock Selfdoc: Self-supervised document representation learning.
\newblock In \emph{Proceedings of the IEEE/CVF Conference on Computer Vision
  and Pattern Recognition}, pages 5652--5660.

\bibitem[{Li et~al.(2025)Li, Xie, Sun, Chen, Qin, Ke, and Yan}]{li2025min}
Shuqi Li, Shufang Xie, Hongda Sun, Yuhan Chen, Tao Qin, Tianjun Ke, and Rui
  Yan. 2025.
\newblock Min: Multi-channel interaction network for drug-target interaction
  with protein distillation.
\newblock \emph{IEEE Transactions on Computational Biology and Bioinformatics}.

\bibitem[{Lin et~al.(2023)Lin, Akin, Rao, Hie, Zhu, Lu, Smetanin, Verkuil,
  Kabeli, Shmueli et~al.}]{lin2023evolutionary}
Zeming Lin, Halil Akin, Roshan Rao, Brian Hie, Zhongkai Zhu, Wenting Lu, Nikita
  Smetanin, Robert Verkuil, Ori Kabeli, Yaniv Shmueli, and 1 others. 2023.
\newblock Evolutionary-scale prediction of atomic-level protein structure with
  a language model.
\newblock \emph{Science}, 379(6637):1123--1130.

\bibitem[{Liu et~al.(2021)Liu, Jiao, Massiah, Yilmaz, and
  Havrylov}]{liu2021trans}
Fangyu Liu, Yunlong Jiao, Jordan Massiah, Emine Yilmaz, and Serhii Havrylov.
  2021.
\newblock Trans-encoder: Unsupervised sentence-pair modelling through self-and
  mutual-distillations.
\newblock In \emph{International Conference on Learning Representations}.

\bibitem[{Liu et~al.(2015)Liu, Sun, Guan, Zheng, and Zhou}]{liu2015improving}
Hui Liu, Jianjiang Sun, Jihong Guan, Jie Zheng, and Shuigeng Zhou. 2015.
\newblock Improving compound--protein interaction prediction by building up
  highly credible negative samples.
\newblock \emph{Bioinformatics}, 31(12):i221--i229.

\bibitem[{Mysinger et~al.(2012)Mysinger, Carchia, Irwin, and
  Shoichet}]{mysinger2012directory}
Michael~M Mysinger, Michael Carchia, John~J Irwin, and Brian~K Shoichet. 2012.
\newblock Directory of useful decoys, enhanced (dud-e): better ligands and
  decoys for better benchmarking.
\newblock \emph{Journal of medicinal chemistry}, 55(14):6582--6594.

\bibitem[{Nga et~al.(2025)Nga, Pham, and Hy}]{nga2025lantern}
Ha~Cong Nga, Phuc Pham, and Truong~Son Hy. 2025.
\newblock Lantern: Leveraging large language models and transformers for
  enhanced molecular interactions.
\newblock \emph{bioRxiv}, pages 2025--02.

\bibitem[{Nguyen et~al.(2021)Nguyen, Le, Quinn, Nguyen, Le, and
  Venkatesh}]{nguyen2021graphdta}
Thin Nguyen, Hang Le, Thomas~P Quinn, Tri Nguyen, Thuc~Duy Le, and Svetha
  Venkatesh. 2021.
\newblock Graphdta: predicting drug--target binding affinity with graph neural
  networks.
\newblock \emph{Bioinformatics}, 37(8):1140--1147.

\bibitem[{Pei et~al.(2023)Pei, Zhang, Zhu, Wu, Gao, Wu, Xia, and
  Yan}]{pei-etal-2023-biot5}
Qizhi Pei, Wei Zhang, Jinhua Zhu, Kehan Wu, Kaiyuan Gao, Lijun Wu, Yingce Xia,
  and Rui Yan. 2023.
\newblock {B}io{T}5: Enriching cross-modal integration in biology with chemical
  knowledge and natural language associations.
\newblock In \emph{Proceedings of the 2023 Conference on Empirical Methods in
  Natural Language Processing}, pages 1102--1123, Singapore. Association for
  Computational Linguistics.

\bibitem[{Peng et~al.(2024)Peng, Liu, Yang, Liu, Bai, Chen, Lu, and
  Nie}]{peng2024bindti}
Lihong Peng, Xin Liu, Long Yang, Longlong Liu, Zongzheng Bai, Min Chen, Xu~Lu,
  and Libo Nie. 2024.
\newblock Bindti: A bi-directional intention network for drug-target
  interaction identification based on attention mechanisms.
\newblock \emph{IEEE Journal of Biomedical and Health Informatics}.

\bibitem[{Rai et~al.(2017)Rai, Brimacombe, Mott, Urban, Hu, Yang, Lee, Cheff,
  Kouznetsova, Benavides et~al.}]{rai2017discovery}
Ganesha Rai, Kyle~R Brimacombe, Bryan~T Mott, Daniel~J Urban, Xin Hu, Shyh-Ming
  Yang, Tobie~D Lee, Dorian~M Cheff, Jennifer Kouznetsova, Gloria~A Benavides,
  and 1 others. 2017.
\newblock Discovery and optimization of potent, cell-active pyrazole-based
  inhibitors of lactate dehydrogenase (ldh).
\newblock \emph{Journal of medicinal chemistry}, 60(22):9184--9204.

\bibitem[{Rogers and Hahn(2010)}]{rogers2010extended}
David Rogers and Mathew Hahn. 2010.
\newblock Extended-connectivity fingerprints.
\newblock \emph{Journal of chemical information and modeling}, 50(5):742--754.

\bibitem[{Rong et~al.(2020)Rong, Bian, Xu, Xie, Wei, Huang, and
  Huang}]{rong2020self}
Yu~Rong, Yatao Bian, Tingyang Xu, Weiyang Xie, Ying Wei, Wenbing Huang, and
  Junzhou Huang. 2020.
\newblock Self-supervised graph transformer on large-scale molecular data.
\newblock \emph{Advances in neural information processing systems},
  33:12559--12571.

\bibitem[{Ross et~al.(2022)Ross, Belgodere, Chenthamarakshan, Padhi, Mroueh,
  and Das}]{ross2022large}
Jerret Ross, Brian Belgodere, Vijil Chenthamarakshan, Inkit Padhi, Youssef
  Mroueh, and Payel Das. 2022.
\newblock Large-scale chemical language representations capture molecular
  structure and properties.
\newblock \emph{Nature Machine Intelligence}, 4(12):1256--1264.

\bibitem[{Schenone et~al.(2013)Schenone, Dan{\v{c}}{\'\i}k, Wagner, and
  Clemons}]{schenone2013target}
Monica Schenone, Vlado Dan{\v{c}}{\'\i}k, Bridget~K Wagner, and Paul~A Clemons.
  2013.
\newblock Target identification and mechanism of action in chemical biology and
  drug discovery.
\newblock \emph{Nature chemical biology}, 9(4):232--240.

\bibitem[{Su et~al.(2023)Su, Han, Zhou, Shan, Zhou, and Yuan}]{su2023saprot}
Jin Su, Chenchen Han, Yuyang Zhou, Junjie Shan, Xibin Zhou, and Fajie Yuan.
  2023.
\newblock Saprot: protein language modeling with structure-aware vocabulary.
\newblock \emph{Advances in neural information processing systems}, pages
  2023--10.

\bibitem[{Su et~al.(2024)Su, Li, Han, Zhou, Shan, Zhou, Ma, OPMC, Ovchinnikov,
  and Yuan}]{su2024saprothub}
Jin Su, Zhikai Li, Chenchen Han, Yuyang Zhou, Junjie Shan, Xibin Zhou, Dacheng
  Ma, The OPMC, Sergey Ovchinnikov, and Fajie Yuan. 2024.
\newblock Saprothub: Making protein modeling accessible to all biologists.
\newblock \emph{bioRxiv}, pages 2024--05.

\bibitem[{Sun et~al.(2024)Sun, Li, Leung, and Hu}]{sun2024ingnn}
Yan Sun, Yan~Yi Li, Carson~K Leung, and Pingzhao Hu. 2024.
\newblock ingnn-dti: prediction of drug--target interaction with interpretable
  nested graph neural network and pretrained molecule models.
\newblock \emph{Bioinformatics}, 40(3):btae135.

\bibitem[{Svensson et~al.(2024)Svensson, Hoedt, Hochreiter, and
  Klambauer}]{svensson2024hyperpcm}
Emma Svensson, Pieter-Jan Hoedt, Sepp Hochreiter, and Günter Klambauer. 2024.
\newblock Hyperpcm: Robust task-conditioned modeling of drug--target
  interactions.
\newblock \emph{Journal of Chemical Information and Modeling},
  64(7):2539--2553.

\bibitem[{Tang et~al.(2014)Tang, Szwajda, Shakyawar, Xu, Hintsanen, Wennerberg,
  and Aittokallio}]{tang2014making}
Jing Tang, Agnieszka Szwajda, Sushil Shakyawar, Tao Xu, Petteri Hintsanen,
  Krister Wennerberg, and Tero Aittokallio. 2014.
\newblock Making sense of large-scale kinase inhibitor bioactivity data sets: a
  comparative and integrative analysis.
\newblock \emph{Journal of chemical information and modeling}, 54(3):735--743.

\bibitem[{Van~Kempen et~al.(2024)Van~Kempen, Kim, Tumescheit, Mirdita, Lee,
  Gilchrist, S{\"o}ding, and Steinegger}]{van2024fast}
Michel Van~Kempen, Stephanie~S Kim, Charlotte Tumescheit, Milot Mirdita,
  Jeongjae Lee, Cameron~LM Gilchrist, Johannes S{\"o}ding, and Martin
  Steinegger. 2024.
\newblock Fast and accurate protein structure search with foldseek.
\newblock \emph{Nature Biotechnology}, 42(2):243--246.

\bibitem[{Varadi et~al.(2022)Varadi, Anyango, Deshpande, Nair, Natassia,
  Yordanova, Yuan, Stroe, Wood, Laydon et~al.}]{varadi2022alphafold}
Mihaly Varadi, Stephen Anyango, Mandar Deshpande, Sreenath Nair, Cindy
  Natassia, Galabina Yordanova, David Yuan, Oana Stroe, Gemma Wood, Agata
  Laydon, and 1 others. 2022.
\newblock Alphafold protein structure database: massively expanding the
  structural coverage of protein-sequence space with high-accuracy models.
\newblock \emph{Nucleic acids research}, 50(D1):D439--D444.

\bibitem[{Vaswani et~al.(2017)Vaswani, Shazeer, Parmar, Uszkoreit, Jones,
  Gomez, Kaiser, and Polosukhin}]{vaswani2017attention}
Ashish Vaswani, Noam Shazeer, Niki Parmar, Jakob Uszkoreit, Llion Jones,
  Aidan~N Gomez, {\L}ukasz Kaiser, and Illia Polosukhin. 2017.
\newblock Attention is all you need.
\newblock \emph{Advances in neural information processing systems}, 30.

\bibitem[{Vig(2019)}]{vig-2019-multiscale}
Jesse Vig. 2019.
\newblock A multiscale visualization of attention in the transformer model.
\newblock In \emph{Proceedings of the 57th Annual Meeting of the Association
  for Computational Linguistics: System Demonstrations}, pages 37--42,
  Florence, Italy. Association for Computational Linguistics.

\bibitem[{Waliany et~al.(2025)Waliany, Lin, and Gainor}]{waliany2025evolution}
Sarah Waliany, Jessica~J Lin, and Justin~F Gainor. 2025.
\newblock Evolution of first versus next-line targeted therapies for metastatic
  non-small cell lung cancer.
\newblock \emph{Trends in Cancer}.

\bibitem[{Weininger(1988)}]{weininger1988smiles}
David Weininger. 1988.
\newblock Smiles, a chemical language and information system. 1. introduction
  to methodology and encoding rules.
\newblock \emph{Journal of chemical information and computer sciences},
  28(1):31--36.

\bibitem[{Weininger et~al.(1989)Weininger, Weininger, and
  Weininger}]{weininger1989smiles}
David Weininger, Arthur Weininger, and Joseph~L Weininger. 1989.
\newblock Smiles. 2. algorithm for generation of unique smiles notation.
\newblock \emph{Journal of chemical information and computer sciences},
  29(2):97--101.

\bibitem[{Wishart et~al.(2008)Wishart, Knox, Guo, Cheng, Shrivastava, Tzur,
  Gautam, and Hassanali}]{wishart2008drugbank}
David~S Wishart, Craig Knox, An~Chi Guo, Dean Cheng, Savita Shrivastava, Dan
  Tzur, Bijaya Gautam, and Murtaza Hassanali. 2008.
\newblock Drugbank: a knowledgebase for drugs, drug actions and drug targets.
\newblock \emph{Nucleic acids research}, 36(suppl\_1):D901--D906.

\bibitem[{Wu et~al.(2022)Wu, Gao, Zeng, Zhang, and Li}]{wu2022bridgedpi}
Yifan Wu, Min Gao, Min Zeng, Jie Zhang, and Min Li. 2022.
\newblock Bridgedpi: a novel graph neural network for predicting drug--protein
  interactions.
\newblock \emph{Bioinformatics}, 38(9):2571--2578.

\bibitem[{Xu et~al.(2023)Xu, Yuan, Miret, and Tang}]{xu2023protst}
Minghao Xu, Xinyu Yuan, Santiago Miret, and Jian Tang. 2023.
\newblock Protst: Multi-modality learning of protein sequences and biomedical
  texts.
\newblock In \emph{International Conference on Machine Learning}, pages
  38749--38767. PMLR.

\bibitem[{Yang et~al.(2021)Yang, Zhong, Zhao, and Chen}]{yang2021ml}
Ziduo Yang, Weihe Zhong, Lu~Zhao, and Calvin Yu-Chian Chen. 2021.
\newblock Ml-dti: mutual learning mechanism for interpretable drug--target
  interaction prediction.
\newblock \emph{The Journal of Physical Chemistry Letters}, 12(17):4247--4261.

\bibitem[{Ying et~al.(2021)Ying, Cai, Luo, Zheng, Ke, He, Shen, and
  Liu}]{ying2021transformers}
Chengxuan Ying, Tianle Cai, Shengjie Luo, Shuxin Zheng, Guolin Ke, Di~He,
  Yanming Shen, and Tie-Yan Liu. 2021.
\newblock Do transformers really perform badly for graph representation?
\newblock \emph{Advances in neural information processing systems},
  34:28877--28888.

\bibitem[{Y{\"u}ksel et~al.(2023)Y{\"u}ksel, Ulusoy, {\"U}nl{\"u}, and
  Do{\u{g}}an}]{yuksel2023selformer}
Atakan Y{\"u}ksel, Erva Ulusoy, Atabey {\"U}nl{\"u}, and Tunca Do{\u{g}}an.
  2023.
\newblock Selformer: molecular representation learning via selfies language
  models.
\newblock \emph{Machine Learning: Science and Technology}, 4(2):025035.

\bibitem[{Zeng et~al.(2024)Zeng, Chen, and Lei}]{zeng2024cat}
Xiaoting Zeng, Weilin Chen, and Baiying Lei. 2024.
\newblock Cat-dti: cross-attention and transformer network with domain
  adaptation for drug-target interaction prediction.
\newblock \emph{BMC bioinformatics}, 25(1):141.

\bibitem[{Zhang et~al.(2024)Zhang, Gong, Pan, Wu, Du, and Hu}]{zhang2024cross}
Hongzhi Zhang, Xiuwen Gong, Shirui Pan, Jia Wu, Bo~Du, and Wenbin Hu. 2024.
\newblock A cross-field fusion strategy for drug-target interaction prediction.
\newblock \emph{arXiv preprint arXiv:2405.14545}.

\bibitem[{Zheng et~al.(2020)Zheng, Li, Chen, Xu, and Yang}]{zheng2020DrugVQA}
Shuangjia Zheng, Yongjian Li, Sheng Chen, Jun Xu, and Yuedong Yang. 2020.
\newblock Predicting drug--protein interaction using quasi-visual question
  answering system.
\newblock \emph{Nature Machine Intelligence}, 2(2):134--140.

\bibitem[{Zitnik et~al.(2018)Zitnik, Sosic, and Leskovec}]{zitnik2018biosnap}
Marinka Zitnik, Rok Sosic, and Jure Leskovec. 2018.
\newblock Biosnap datasets: Stanford biomedical network dataset collection.
\newblock \emph{Note: http://snap. stanford. edu/biodata Cited by}, 5(1).

\end{thebibliography}

\appendix

\section{Appendix}
\label{sec:appendix}

\subsection{Hyperparameter of FusionDTI}
\label{subsec:hyperparameter}

FusionDTI is implemented in Python 3.8 and the PyTorch framework (1.12.1)\footnote{\url{https://pytorch.org/}}. The computing device we use is the NVIDIA GeForce RTX 3090. In the "Experimental Setup and Results" section, we only present experiment results based on the BindingDB dataset, as the performance trends are identical to the BioSNAP dataset and the Human dataset. Table~\ref{FusionDTI-hyperparams} shows the parameters of the FusionDTI model and Table~\ref{Notation_new} lists the notations used in this paper with descriptions. 

\begin{table*}[hbt!]
\centering
\setlength{\tabcolsep}{1mm}{
\begin{tabular}{lll}
\toprule
Module & Hyperparameter & Value \\ \midrule
Mini-batch          & Batch size & 64 (options: 64, 128)\\
Drug Encoder    & PLM & \href{https://huggingface.co/HUBioDataLab/SELFormer}{HUBioDataLab/SELFormer} \\   
Protein Encoder & PLM & \href{https://huggingface.co/westlake-repl/SaProt_650M_AF2}{westlake-repl/SaProt\_650M\_AF2} \\
BAN             & Heads of bilinear attention & 3 \\
                & Bilinear embedding size & 512 (options: 32, 64, 128, 256, 512, 768)\\
                & Sum pooling window size & 2 \\
CAN           & Attention heads & 8 \\
              & Hidden dimension & 512 (options: 32, 64, 128, 256, 512, 768)\\
              & Integration strategies & Mean pooling (options: Mean pooling, CLS)\\
              & Group size & 1 (options: from 1 to 512) \\
MLP & Hidden layer sizes & (1024, 512, 256) \\
    & Activation & Relu (options: Tanh, Relu) \\
    & Solver & AdamW \\ & & (options: AdamW, Adam, RMSprop, Adadelta, LBFGS) \\
    & Learning rate scheduler & CosineAnnealingLR \\ & &  (options: CosineAnnealingLR, StepLR, ExponentialLR) \\
    & Initial learning rate & 1e-4 (options: from 1e-3 to 1e-6) \\
    & Maximum epoch & 200 \\
\bottomrule
\end{tabular}}
\caption{\centering\label{FusionDTI-hyperparams} Configuration Parameters}

\end{table*}

\begin{table*}[hbt!]
\centering
\setlength{\tabcolsep}{1mm}{
\begin{tabular}{ll}
\toprule
Notations & Description \\ \midrule
$\mathbf{D} $ & Drug feature \\
$\mathbf{P} $ & Target feature \\
$\mathbf{q} \in \mathbb{R}^K$ &  weight vector for bilinear transformation\\
$Att \in \mathbb{R}^{\rho \times \phi}$ & Bilinear attention maps in BAN \\
$\phi, \rho$ & The feature dimensions of drug and protein embeddings \\
$\mathbf{U} \in \mathbb{R}^{n \times K}$ & Transformation matrix for drug features \\
$\mathbf{V} \in \mathbb{R}^{m \times K}$ & Transformation matrix for target features \\
$m, n$ & The maximum allowed sequence length of drugs and protein \\
$\mathbf{g} $ & The number of tokens per group\\
$\mathbf{D}^{*} \in \mathbb{R}^{m \times h}$ & Fused drug representations in token-level interaction \\
$\mathbf{P}^{*} \in \mathbb{R}^{n \times h}$ & Fused target representations in token-level interaction \\
$\mathbf{Q}_d, \mathbf{K}_d, \mathbf{V}_d \in \mathbb{R}^{m \times h}$ & Queries, keys, and values for the drug in token-level interaction \\
$\mathbf{Q}_p, \mathbf{K}_p, \mathbf{V}_p \in \mathbb{R}^{n \times h}$ & Queries, keys, and values for target in token-level interaction \\
$\mathbf{W}_q^d, \mathbf{W}_k^d, \mathbf{W}_v^d \in \mathbb{R}^{H \times h}$ & Projection matrices for drug queries, keys, and values \\
$\mathbf{W}_q^p, \mathbf{W}_k^p, \mathbf{W}_v^p \in \mathbb{R}^{h \times h}$ & Projection matrices for target queries, keys, and values \\
$\mathbf{F} $ & drug-target joint representation \\
$p \in [0, 1]$ & output interaction probability \\
$H$ & Number of attention heads in token-level interaction \\
$h$ & Hidden dimension in token-level interaction \\
\bottomrule
\end{tabular}}
\caption{\centering\label{Notation_new} Notations and Descriptions}
\end{table*}

\subsection{Dataset Sources}
\label{subsec:datasets}

All the data used in this paper are from public sources. The statistics of the experimental datasets are presented in Table~\ref{dataset}.  

\begin{table}[hbt!]
\centering
\footnotesize
\begin{tabularx}{\columnwidth}{XXXX}
\toprule
Dataset &  Drugs &  Proteins &  Interactions   \\ \midrule
BindingDB  & 14,643 & 2,623 & 49,199   \\
BioSNAP & 4,510 & 2,181 & 27,464\\
Human & 2,726 & 2,001 & 6,728\\
DAVIS & 68 & 442 & 30,056\\
KIBA & 2,068 & 229 & 118,254\\ 
DUD-E & 1,200,966 & 102 & 1,434,019\\
\bottomrule
\end{tabularx}
\caption{\label{dataset}Dataset Statistics.}
\end{table}

\begin{enumerate}
    \item The BindingDB~\citep{gilson2016bindingdb} dataset is a web-accessible database of experimentally validated binding affinities, focusing primarily on the interactions of small drug-like molecules and proteins. The BindingDB source is found at \url{https://www.bindingdb.org/bind/index.jsp}.
    \item The BioSNAP~\citep{zitnik2018biosnap} dataset is created from the DrugBank database~\citep{wishart2008drugbank}. It is a balanced dataset with validated positive interactions and an equal number of negative samples randomly obtained from unseen pairs. The BioSNAP source is found at \url{https://github.com/kexinhuang12345/MolTrans}.
    \item The Human~\citep{liu2015improving,chen2020transformercpi} dataset includes highly credible negative samples. The balanced version of the Human dataset contains the same number of positive and negative samples. The Human source is found at \url{https://github.com/lifanchen-simm/transformerCPI}.
    \item The DAVIS~\citep{davis2011comprehensive} dataset provides continuous binding affinity measurements (\textit{K$_d$} values) between kinase inhibitors and proteins. It is commonly used for regression-based drug–target interaction (DTI) prediction tasks. The DAVIS source is available at \url{https://tdcommons.ai/multi_pred_tasks/dti/}.
    \item The KIBA~\citep{tang2014making} dataset integrates multiple bioactivity measures to provide a unified KIBA score for kinase–inhibitor pairs. It is widely adopted in benchmark studies for affinity prediction. The KIBA source is available at \url{https://tdcommons.ai/multi_pred_tasks/dti/}.
    \item The DUD-E~\citep{mysinger2012directory} (Directory of Useful Decoys, Enhanced) dataset is a large-scale benchmark set for virtual screening, containing active compounds and challenging decoys for various protein targets. The DUD-E source is found at \url{http://dude.docking.org/}.
\end{enumerate}

\subsection{How to Obtain the Structure-aware (SA) Sequence of a Protein and the SELFIES of a Drug?}
\label{subsec:obtain_sequence}

To obtain the SA sequence of a protein, the first step is to obtain Uniprot IDs from the \href{https://www.uniprot.org}{UniProt website} using information such as the amino acid sequences or protein names, and then save these IDs in a comma-delimited text file. Subsequently, we use the UniProt IDs to fetch the relevant 3D structure file (.cif) from \href{https://alphafold.ebi.ac.uk}{AlphafoldDB}~\citep{varadi2022alphafold} using Foldseek. The SA vocabulary of the protein can then be generated from this 3D structure file.

For drugs, the SELFIES could be derived from SMILES strings. This conversion requires specific Python packages, and upon installation, the SELFIES strings can be generated through appropriate scripts. Please refer to our submission file for detailed procedures, including the necessary code.

% Notably, we provide the generation code for SA vocabulary and SELFIES in our \href{https://github.com/ZhaohanM/FusionDTI}{GitHub}.
Notably, our submission of supplementary material contains step-by-step descriptions and code for generating the SA sequences and SELFIES.

\subsection{Baselines}
\label{subsec:baseline}

We compare the performance of FusionDTI with the following eight models on the DTI task.

\paragraph{Baselines on BindingDB, BioSNAP, and Human.}
\begin{enumerate}
    \item Support Vector Machine~\citep{cortes1995support} on the concatenated fingerprint ECFP4~\citep{rogers2010extended} (extended connectivity fingerprint, up to four bonds) and PSC~\citep{cao2013propy} (pseudo-amino acid composition) features.
    
    \item Random Forest~\citep{ho1995random} on the concatenated fingerprint ECFP4 and PSC features.
    
    \item DeepConv-DTI~\citep{lee2019deepconv} uses a fully connected neural network to encode the ECFP4 drug fingerprint and a CNN along with a global max-pooling layer to extract features from the protein sequences. Then the drug and protein features are concatenated and fed into a fully connected neural network for the final prediction.

    \item GraphDTA~\citep{nguyen2021graphdta} uses GNN for the encoding of drug molecular graphs, and a CNN is used for the encoding of the protein sequences. The derived vectors of the drug and protein representations are directly concatenated for interaction prediction.

    \item MolTrans~\citep{huang2021moltrans} uses a transformer architecture to encode the drugs and proteins. Then a CNN-based fusion module is adapted to capture DTI interactions.

    \item DrugBAN~\citep{bai2023interpretable} use a Graph Convolution Network and 1D CNN to encode the drug and protein sequences. Then a bilinear attention network~\citep{kim2018bilinear} is adopted to learn pairwise interactions between the drug and protein. The resulting joint representation is decoded by a fully connected neural network.

    \item BioT5~\citep{pei-etal-2023-biot5} is a cross-modelling model in biology with chemical knowledge and natural language associations.

    \item SiamDTI~\citep{zhang2024cross} is a double-channel network structure to acquire local and global protein information for cross-field supervised learning.
\end{enumerate}

\paragraph{Baselines on DAVIS and KIBA.}
\begin{enumerate}
    \setcounter{enumi}{8}
    
    \item \textbf{ML-DTI}~\citep{yang2021ml} combines molecular fingerprints with physicochemical descriptors and applies MLPs for regression.

    \item \textbf{DGraphDTA (Alphafold2)}~\citep{wu2022bridgedpi} integrates protein 3D structural data (from AlphaFold2) with drug graphs through a dual-graph encoding strategy.

    \item \textbf{iNGNN-DTI}~\citep{sun2024ingnn} introduces an interpretable graph neural network with attention-based gating mechanisms for drug–target regression tasks.
    
    \item \textbf{MIN}~\citep{li2025min} uses a hierarchical multi-channel network that combines structure-aware and structure-agnostic representations with interpretable attention mechanisms.

    \item \textbf{LANTERN}~\citep{nga2025lantern} is a versatile deep learning framework that integrates PLMs with transformer-based fusion to deliver structure-free prediction of diverse molecular interactions including DTI, DDI, and PPI.
    
\end{enumerate}

\paragraph{Baselines on DUD-E.}
\begin{enumerate}
    \setcounter{enumi}{12}
    \item \textbf{DrugVQA}~\citep{zheng2020DrugVQA} formulates DTI prediction as a visual question answering task over molecular structures and protein sequences.

    \item \textbf{DrugClip}~\citep{gao2023drugclip} adapts a contrastive pretraining framework, aligning drug molecules and protein embeddings using a CLIP-style architecture.

    \item \textbf{HyperPCM}~\citep{svensson2024hyperpcm} utilises hyperbolic protein–compound matching for robust generalisation in few-shot virtual screening scenarios.

    \item \textbf{MIN}~\citep{li2025min} introduces multi-instance networks to model DTI at the binding site level using hierarchical attention.
\end{enumerate}

\subsection{Ablation Study}
\label{subsec:ablation_study}

In Table~\ref{tab:Aggregation}, we compare the performance of two aggregation strategies within the CAN module. The pooling strategy outperforms the CLS-based aggregation, achieving an AUC and AUPRC of 0.989 and 0.990, respectively. This comparison highlights the superior effectiveness of the pooling in aggregating contextual information. Thus, the integration of a CAN module, particularly employing a pooling aggregation strategy, is shown to be essential for making confident and accurate predictions.

\subsection{Evaluation of PLMs Encoding}
\label{subsec:PLMs}

\begin{figure*}[hbpt!]
    \begin{minipage}{0.49\textwidth}
        \centering
         \includegraphics[width=\linewidth]{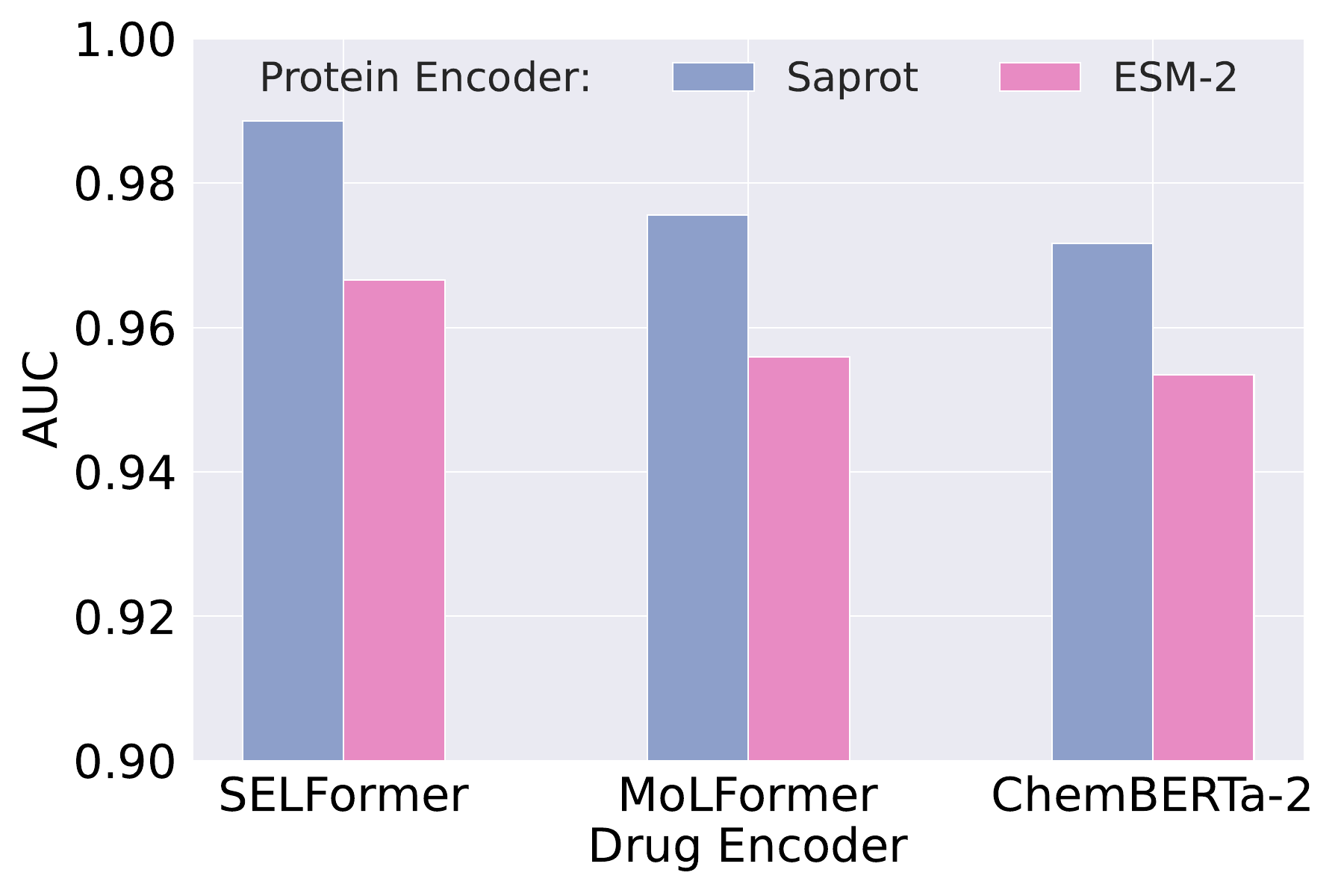}
         \caption{Performance comparison of protein encoders on the BindingDB dataset.}
        \label{fig:PLMs_comparison_protein}
    \end{minipage}%
    \hfill
    \begin{minipage}{0.49\textwidth}
        \centering
        \includegraphics[width=\textwidth]{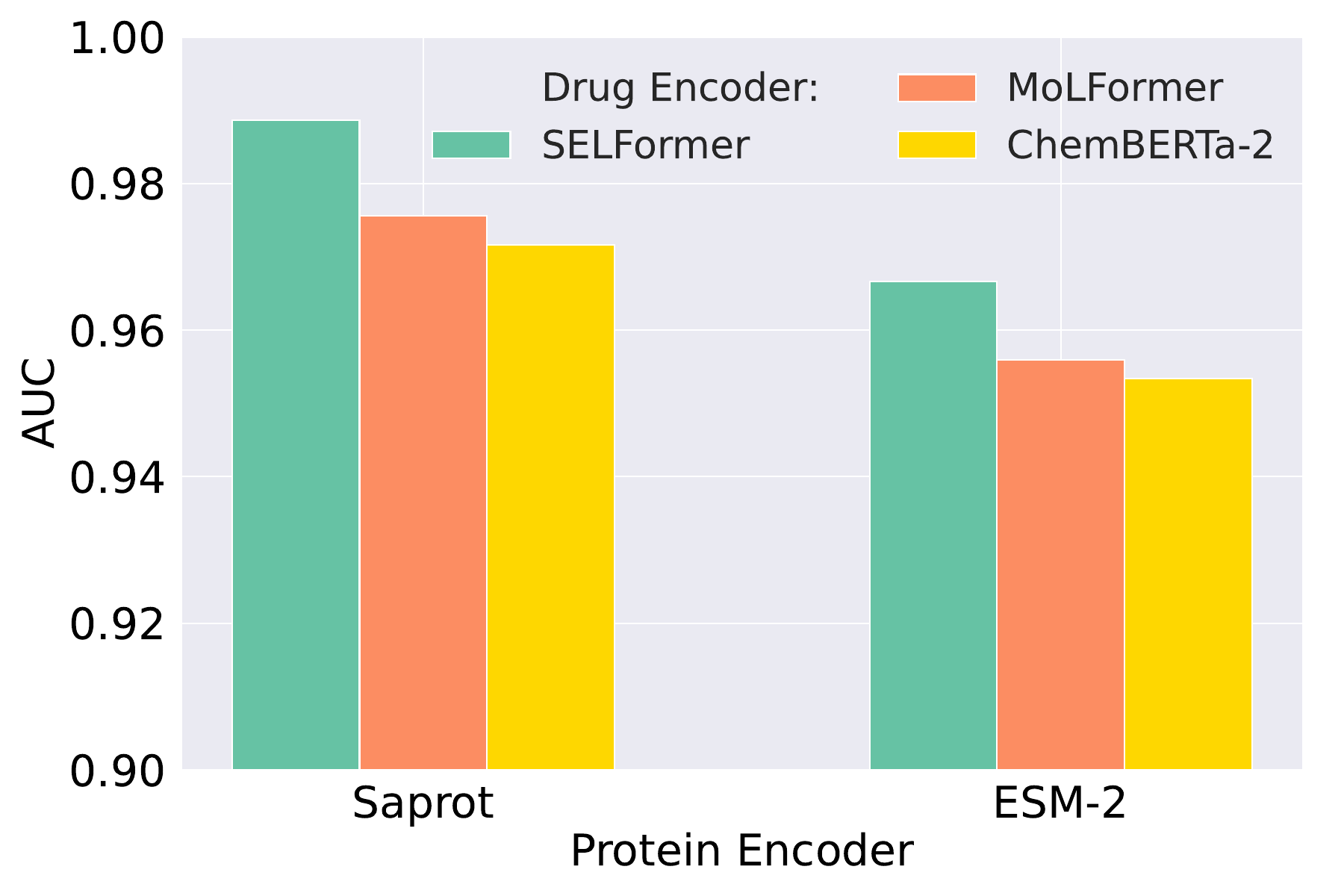} 
        \caption{Performance comparison of drug encoders on the BindingDB dataset.}
        \label{fig:PLMs_comparison_drug}
    \end{minipage}
\end{figure*}

The protein encoder and drug encoder are fundamental for the token-level fusion of representations, as these encoders are responsible for generating fine-grained representations to better explore interaction information. Our proposed model employs two PLMs encoding two biomedical entities: the drug and protein, respectively. In terms of the protein encoders, Figure~\ref{fig:PLMs_comparison_protein} compares the the performance of the two protein encoders (SaProt~\citep{su2023saprot} and ESM-2~\citep{lin2023evolutionary}) in combination with three different drug encoders: ChemBERTa-2~\citep{ahmad2022chemberta}, SELFormer~\citep{yuksel2023selformer} and MoLFormer~\citep{ross2022large}. From the figure, we find that SaProt consistently outperforms ESM-2 when combined with all three drug encoders. As can be seen in Figure~\ref{fig:PLMs_comparison_drug}, SELFormer achieves the best performance in encoding the drug sequences among the three advanced drug encoders. Notably, the top-performing combination is SaProt and SELFormer, hence our proposed FusionDTI uses them as drug and protein encoders.

\subsection{Efficiency Analysis}
\label{subsec:Efficiency_Analysis}

\begin{figure}[t]
    \centering
    \scriptsize
    \resizebox{\columnwidth}{!}{%
    \includegraphics{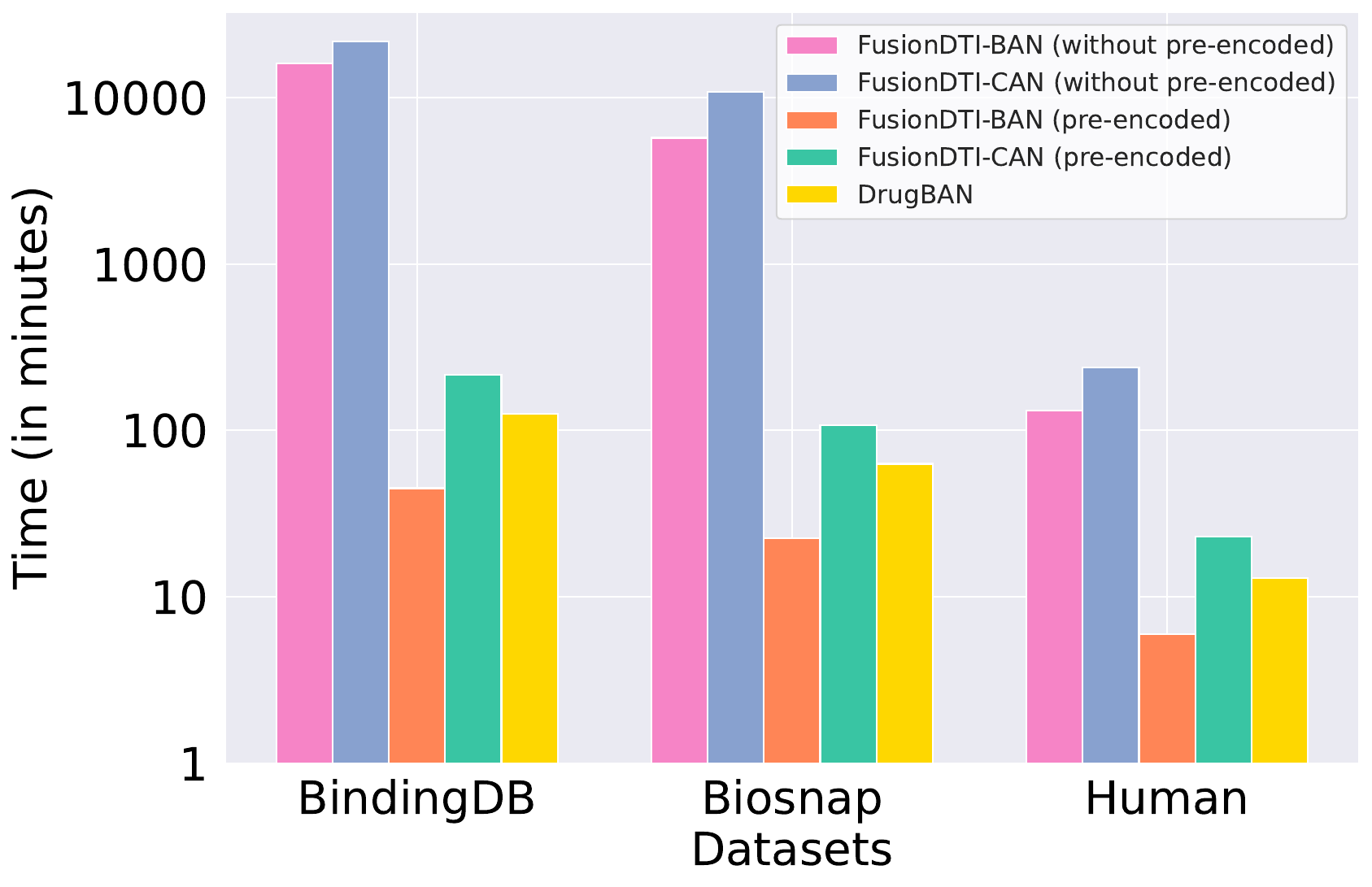}}
\caption{Time comparison on the BindingDB, Human and BioSNAP datasets.}
\label{fig:Efficiency_Analysis}
\end{figure}

\begin{table}[t]
        \centering
        \footnotesize
        \begin{tabularx}{\columnwidth}{XXXX}
            \toprule
             Aggregation & AUC  & AUPRC & Accuracy\\
            \midrule
             CLS  & 0.982 & 0.983 & 0.956\\
             Pooling  & 0.989 & 0.990 & 0.961\\
            \bottomrule
        \end{tabularx}
        \caption{\label{tab:Aggregation}Comparison of aggregation strategies for FusionDTI-CAN on the BindingDB dataset.}
\end{table}

Efficiency in computational models is crucial, particularly when handling large-scale and extensive datasets in drug discovery. Our proposed model stores drug representations and target representations in memory for later online training. As evidenced by Figure~\ref{fig:Efficiency_Analysis}, FusionDTI-CAN and FusionDTI-BAN with pre-encoded representations process the BindingDB dataset much faster than the non-pre-coded models, approximately 45 minutes and 220 minutes, respectively. This stark difference highlights the advantage of pre-encoding, which eliminates the need for real-time data processing and accelerates the overall throughput. While FusionDTI-BAN and DrugBAN have the same fusion module, the pre-encoded FusionDTI-BAN runs faster and predicts more accurately, as shown in Table~\ref{tab:in_domain}. In addition, FusionDTI-BAN runs faster than FusionDTI-CAN, indicating that the BAN fusion module is more efficient. Ultimately, FusionDTI-BAN with pre-encoded data stands out as a highly efficient approach, offering substantial benefits in scenarios where large-scale data exists. 

% We further analyse the time complexity in Appendix~\ref{subsec:time_complexity}.

\subsection{Time Complexity Analysis}
\label{subsec:time_complexity}

\begin{table}[hbpt!]
\vspace{1em} 
\centering
\begin{tabular}{ccc}
\toprule
Fusion module & Complexity (O) & Parameters \\ 
\midrule
BAN & \(O(\rho \cdot \phi \cdot K)\) & 790k \\ 
CAN & \(O(m \cdot n \cdot h)\) & 1572k \\ 
\bottomrule
\end{tabular} 
\caption{\label{tab:complexity}Time complexity and parameters comparison of BAN and CAN.}
\end{table}

\begin{table*}[t]
    \centering
    \footnotesize
    \begin{tabularx}{\textwidth}{>{\centering\arraybackslash}p{0.2\textwidth} >{\centering\arraybackslash}X}
        \toprule
        \textbf{Drug-Target (Ligand - PDB ID)} & \textbf{Predicted Binding Residues} \\
        \midrule
        VGH - 2YFX & \textbf{Glu113}, \textbf{Val46}, \textbf{Gly117}, \textbf{Met115}, \textbf{Asp186}, Arg125, Lys225, Gln50, Ala190, Pro319\\
        \midrule
        C6F - 6JQR & \textbf{Tyr126}, \textbf{Asp209}, \textbf{Ala72}, \textbf{Glu208}, \textbf{Glu197}, Leu219, Pro163, Gln97,  Val225, His151 \\
        \midrule
        5P8 - 4CLI &  \textbf{GLu113}, \textbf{Leu172}, \textbf{Gly118}, \textbf{Ala64}, \textbf{Asp186}, Ala150, Ile99, Pro290, Ala312, Glu316 \\
        \midrule
        0WM - 4G5J & \textbf{His296}, \textbf{Pro102}, \textbf{Pro156}, \textbf{Met295}, \textbf{Asn116}, Ser92, Thr217, Lys237, His143, Trp188 \\
        \midrule
        YY3 - 6LUD & \textbf{Phe102}, \textbf{Leu151}, \textbf{Met1100}, \textbf{Lys52}, \textbf{Glu111}, Ile22, Pro60, Ala129, Val141, Gly42\\
        \midrule
        AQ4 - 1M17 & \textbf{Leu155}, \textbf{Leu99}, \textbf{Met104}, \textbf{Phe106}, \textbf{Thr165}, Asp111, Lys171, Trp209, Ala61, Asp280 \\
        \midrule
        YMX - 5FTO &  \textbf{Asn162}, \textbf{Gly110}, \textbf{Phe35}, \textbf{Glu118}, \textbf{Val38}, His155, ALa197, Met46, Leu112, Asp280\\
        \midrule
        1C9 - 4I23 &  \textbf{Ala50}, \textbf{Leu95}, \textbf{Met100}, \textbf{Pro101}, \textbf{Glu69}, Thr247, Tyr120, His177, Pro221, Val49\\
        \midrule
        VGH - 2XP2 &  \textbf{Leu172}, \textbf{Gly185}, \textbf{Ala116}, \textbf{Lys66}, \textbf{Asp119}, Pro58, Met82, Pro131, Ala167, Val27 \\
        \midrule
        EMH - 3AOX &  \textbf{Glu143}, \textbf{Leu55}, \textbf{Gly56}, \textbf{Val113}, \textbf{Met132}, Glu91, Leu157, Val44, Ala59, Ile166\\
        \bottomrule
    \end{tabularx}
    \caption{Predicted binding sites for DTI in NSCLC. \textbf{Bold} residues are supported by the PDB database, while others remain unverified.}
    \label{tab:nsclc_dti}
\end{table*}

The feature dimensions of the representations generated by different PLM encoders are fixed, but the size of the feature dimensions may not be the same. Therefore, in order to fuse protein and drug representations, we use two linear layers to keep the representations' feature dimension equal to the token length (512).

The time complexity of BAN depends on the computation of bilinear interaction maps. The bilinear attention involves a Hadamard product and further matrix operations as given in Equation~(\ref{eq:att_ban}). The computation of \( U^TP \) and \( V^TD \) requires \( O(N \cdot \rho \cdot K) \) and \( O(M \cdot \phi \cdot K) \) operations, respectively. Here, $ K $ denotes the dimensionality of the transformation, which is the rank of the feature space to which the protein and drug features are projected. When the token length is equal to the feature dimension and the dimensions of transformation are two times either, the overall time complexity is \( O(\rho \cdot \phi \cdot K) \).

For the token-level interaction in the DTI task, the time complexity is also markedly influenced by the attention mechanisms. It also satisfies the condition that the token length is equal to the feature dimension of the drug and protein. With multi-head attention heads (\( H = 8 \)), the complexity for computing the queries, keys, and values in the Equation~(\ref{eq:qkv_drug}) and~(\ref{eq:qkv_protein}), as well as the softmax attention weights, is given by \( O(H \cdot n \cdot m \cdot h) \), where \( m and n \) represents the token lengths for the drug and protein, respectively, and \( h \) is the hidden dimension. Since each head contributes its own set of computations and the attention mechanism operates over all tokens, the \( m \cdot n \) term (stemming from the softmax operation across the token length) becomes significant. This leads to a total time complexity of \( O(m \cdot n \cdot h) \) per batch for the attention mechanism. 

From the above analysis of the time complexity of the two fusion strategies, the time complexity of CAN is lower than BAN in the case of the same input protein and drug features. BAN is markedly affected by the transformation dimension $ K $. When the \( K \) is larger than the token and feature dimension, the time complexity of BAN is higher than CAN. However, we observe that the number of parameters in BAN is smaller than that of CAN via the PyTorch package, as shown in Table~\ref{tab:complexity}.

\subsection{Case Study}
\label{subsec:case_study}

The top three predictions (PDB ID: 6QL2~\cite{kazokaite2019engineered}, 5W8L~\cite{rai2017discovery} and 4N6H~\cite{fenalti2014molecular}) of the co-crystalised ligands are derived from the Protein Data Bank (PDB)~\cite{berman2007worldwide}. Following the setup of the DrugBAN case study, we only chose X-ray structures with a resolution greater than 2.5 Å corresponding to human proteins. In addition, the co-crystalised ligands are required to have pIC$_{50}$ $\le$ 100 nM and are not part of the training dataset. 

To further DTI in non-small cell lung cancer (NSCLC), we identify ten additional drug-protein pairs from PDB. The selected targets—Epidermal Growth Factor Receptor (EGFR), Anaplastic Lymphoma Kinase (ALK), and ROS1—are well-established oncogenic drivers in NSCLC~\citep{waliany2025evolution}. The corresponding inhibitors, including Erlotinib, Gefitinib, Osimertinib, Crizotinib, and Lorlatinib, exhibit high binding affinities~\citep{herrera2023targeted}. Table~\ref{tab:nsclc_dti} presents the predicted binding residues for these interactions, with bolded residues supported by experimental PDB data, while others remain unverified.

\subsection{Performance Comparison}
\label{subsec:performance_comparison}

Tables~\ref{tab:in_domain_full} and~\ref{tab:cross_domain_full} provide a detailed performance evaluation of FusionDTI and baseline models across both in-domain and cross-domain settings. To ensure a comprehensive assessment, we report multiple evaluation metrics, including AUROC and AUPRC as primary indicators, alongside F1-score, Sensitivity, Specificity, and Matthews Correlation Coefficient (MCC). These additional metrics offer deeper insights into model performance across different classification aspects.

In addition, Tables~\ref{tab:davis_full}, \ref{tab:kiba_full}, and~\ref{tab:dude} present results on three benchmark datasets: DAVIS~\citep{davis2011comprehensive}, KIBA~\citep{tang2014making}, and DUD-E~\citep{mysinger2012directory}. Each table compares FusionDTI with strong task-specific baselines under standard evaluation metrics for their respective tasks, further demonstrating the robustness and adaptability of our model.

\begin{table*}[hbpt!]
\centering
\resizebox{\textwidth}{!}{
\huge
\begin{tabular}{c ccccccc}
\toprule
Model & AUC & AUPR & Accuracy & F1 & Sensitivity & Specificity & MCC \\
\midrule
\multicolumn{8}{c}{BindingDB} \\
SVM & 0.939$\pm$0.001 & 0.928$\pm$0.002 & 0.825$\pm$0.004 & 0.821$\pm$0.004 & 0.810$\pm$0.010 & 0.840$\pm$0.007 & 0.700$\pm$0.012 \\
RF & 0.942$\pm$0.011 & 0.921$\pm$0.016 & 0.880$\pm$0.012 & 0.875$\pm$0.012 & 0.870$\pm$0.015 & 0.890$\pm$0.010 & 0.815$\pm$0.009 \\
DeepConv-DTI & 0.945$\pm$0.002 & 0.925$\pm$0.005 & 0.882$\pm$0.007 & 0.878$\pm$0.008 & 0.870$\pm$0.011 & 0.885$\pm$0.010 & 0.818$\pm$0.013 \\
GraphDTA & 0.951$\pm$0.002 & 0.934$\pm$0.002 & 0.888$\pm$0.005 & 0.884$\pm$0.005 & 0.880$\pm$0.006 & 0.890$\pm$0.004 & 0.825$\pm$0.008 \\
MolTrans & 0.952$\pm$0.002 & 0.936$\pm$0.001 & 0.887$\pm$0.006 & 0.882$\pm$0.006 & 0.875$\pm$0.009 & 0.890$\pm$0.007 & 0.820$\pm$0.010 \\
DrugBAN & 0.960$\pm$0.001 & 0.948$\pm$0.002 & 0.906$\pm$0.004 & 0.901$\pm$0.004 & 0.900$\pm$0.008 & 0.908$\pm$0.004 & 0.872$\pm$0.005 \\
SiamDTI & 0.961$\pm$0.002 & 0.945$\pm$0.002 & 0.890$\pm$0.006 & 0.886$\pm$0.006 & 0.880$\pm$0.007 & 0.895$\pm$0.005 & 0.830$\pm$0.006 \\
BioT5 & 0.963$\pm$0.001 & 0.952$\pm$0.001 & 0.907$\pm$0.003 & 0.905$\pm$0.003 & 0.900$\pm$0.004 & 0.910$\pm$0.003 & 0.850$\pm$0.005 \\
\bottomrule
FusionDTI-BAN & \underline{0.975$\pm$0.002} & \underline{0.976$\pm$0.002} & \underline{0.933$\pm$0.003} & \underline{0.934$\pm$0.002} & \underline{0.932$\pm$0.004} & \underline{0.935$\pm$0.003} & \underline{0.900$\pm$0.003} \\
FusionDTI-CAN & \textbf{0.989$\pm$0.002} & \textbf{0.990$\pm$0.002} & \textbf{0.961$\pm$0.002} & \textbf{0.963$\pm$0.012} & \textbf{0.954$\pm$0.003} & \textbf{0.955$\pm$0.012} & \textbf{0.925$\pm$0.023} \\
\midrule
\multicolumn{8}{c}{BioSNAP} \\
SVM & 0.862$\pm$0.007 & 0.864$\pm$0.004 & 0.777$\pm$0.011 & 0.773$\pm$0.011 & 0.760$\pm$0.015 & 0.780$\pm$0.008 & 0.690$\pm$0.013 \\
RF & 0.860$\pm$0.005 & 0.886$\pm$0.005 & 0.804$\pm$0.005 & 0.800$\pm$0.005 & 0.795$\pm$0.008 & 0.810$\pm$0.007 & 0.715$\pm$0.006 \\
DeepConv-DTI & 0.886$\pm$0.006 & 0.890$\pm$0.006 & 0.805$\pm$0.009 & 0.801$\pm$0.009 & 0.800$\pm$0.013 & 0.810$\pm$0.010 & 0.718$\pm$0.012 \\
GraphDTA & 0.887$\pm$0.008 & 0.890$\pm$0.007 & 0.800$\pm$0.007 & 0.796$\pm$0.007 & 0.790$\pm$0.010 & 0.810$\pm$0.009 & 0.712$\pm$0.009 \\
MolTrans & 0.895$\pm$0.004 & 0.897$\pm$0.005 & 0.825$\pm$0.010 & 0.820$\pm$0.010 & 0.815$\pm$0.013 & 0.830$\pm$0.012 & 0.730$\pm$0.011 \\
DrugBAN & 0.903$\pm$0.005 & 0.902$\pm$0.004 & 0.834$\pm$0.008 & 0.830$\pm$0.009 & 0.820$\pm$0.021 & 0.847$\pm$0.010 & 0.719$\pm$0.007 \\
SiamDTI & 0.912$\pm$0.005 & 0.910$\pm$0.003 & 0.855$\pm$0.004 & 0.852$\pm$0.004 & 0.850$\pm$0.006 & 0.860$\pm$0.004 & 0.740$\pm$0.006 \\
BioT5 & \underline{0.937$\pm$0.001} & \underline{0.937$\pm$0.004} & \underline{0.874$\pm$0.001} & \underline{0.870$\pm$0.001} & \underline{0.865$\pm$0.002} & \underline{0.880$\pm$0.003} & \underline{0.765$\pm$0.004} \\
\bottomrule
FusionDTI-BAN & 0.923$\pm$0.002 & 0.921$\pm$0.002 & 0.856$\pm$0.001 & 0.857$\pm$0.001 & 0.854$\pm$0.002 & 0.858$\pm$0.002 & 0.724$\pm$0.001 \\
FusionDTI-CAN & \textbf{0.951$\pm$0.002} & \textbf{0.951$\pm$0.002} & \textbf{0.889$\pm$0.002} & \textbf{0.890$\pm$0.002} & \textbf{0.888$\pm$0.003} & \textbf{0.891$\pm$0.002} & \textbf{0.778$\pm$0.002} \\
\midrule
\multicolumn{8}{c}{Human} \\
SVM & 0.940$\pm$0.006 & 0.920$\pm$0.009 & 0.895$\pm$0.010 & 0.892$\pm$0.011 & 0.880$\pm$0.015 & 0.910$\pm$0.009 & 0.800$\pm$0.012 \\
RF & 0.952$\pm$0.011 & 0.953$\pm$0.010 & 0.920$\pm$0.012 & 0.915$\pm$0.013 & 0.910$\pm$0.017 & 0.930$\pm$0.014 & 0.820$\pm$0.009 \\
DeepConv-DTI & 0.980$\pm$0.002 & 0.981$\pm$0.002 & 0.927$\pm$0.007 & 0.923$\pm$0.006 & 0.920$\pm$0.009 & 0.930$\pm$0.008 & 0.860$\pm$0.010 \\
GraphDTA & 0.981$\pm$0.001 & 0.982$\pm$0.002 & 0.930$\pm$0.008 & 0.925$\pm$0.008 & 0.920$\pm$0.011 & 0.935$\pm$0.009 & 0.870$\pm$0.009 \\
MolTrans & 0.980$\pm$0.002 & 0.978$\pm$0.003 & 0.925$\pm$0.011 & 0.920$\pm$0.012 & 0.915$\pm$0.016 & 0.930$\pm$0.013 & 0.855$\pm$0.010 \\
DrugBAN & 0.982$\pm$0.002 & 0.980$\pm$0.003 & 0.930$\pm$0.004 & 0.903$\pm$0.003 & 0.900$\pm$0.005 & 0.908$\pm$0.004 & 0.810$\pm$0.004 \\
SiamDTI & 0.970$\pm$0.002 & 0.969$\pm$0.003 & 0.920$\pm$0.006 & 0.915$\pm$0.006 & 0.910$\pm$0.008 & 0.925$\pm$0.007 & 0.840$\pm$0.009 \\
BioT5 & \underline{0.989$\pm$0.001} & \underline{0.985$\pm$0.002} & \underline{0.939$\pm$0.008} & \underline{0.937$\pm$0.004} & \underline{0.929$\pm$0.010} & \underline{0.941$\pm$0.004} & \underline{0.892$\pm$0.006} \\
\bottomrule
FusionDTI-BAN & 0.984$\pm$0.002 & 0.984$\pm$0.003 & 0.938$\pm$0.003 & 0.934$\pm$0.002 & 0.927$\pm$0.004 & 0.931$\pm$0.003 & 0.870$\pm$0.003 \\
FusionDTI-CAN & \textbf{0.991$\pm$0.002} & \textbf{0.989$\pm$0.002} & \textbf{0.947$\pm$0.002} & \textbf{0.948$\pm$0.002} & \textbf{0.955$\pm$0.033} & \textbf{0.950$\pm$0.031} & \textbf{0.905$\pm$0.045} \\
\bottomrule
\end{tabular}}
\caption{\label{tab:in_domain_full}In-domain performance comparison of FusionDTI and the baselines on the BindingDB, Human and BioSNAP datasets (\textbf{Best}, \underline{Second Best}).}
\end{table*}

\begin{table*}[hbpt!]
\centering
\resizebox{\textwidth}{!}{
\huge
\begin{tabular}{c ccccccc}
\toprule
Model & AUC & AUPR & Accuracy & F1 & Sensitivity & Specificity & MCC \\
\midrule
\multicolumn{8}{c}{BindingDB} \\
SVM & 0.490$\pm$0.015 & 0.460$\pm$0.001 & 0.531$\pm$0.009 & 0.521$\pm$0.010 & 0.508$\pm$0.015 & 0.548$\pm$0.011 & 0.150$\pm$0.012 \\
RF & 0.493$\pm$0.021 & 0.468$\pm$0.023 & 0.535$\pm$0.012 & 0.525$\pm$0.013 & 0.512$\pm$0.020 & 0.550$\pm$0.014 & 0.162$\pm$0.015 \\
GraphDTA & 0.536$\pm$0.015 & 0.496$\pm$0.029 & 0.472$\pm$0.009 & 0.462$\pm$0.008 & 0.460$\pm$0.014 & 0.478$\pm$0.011 & 0.100$\pm$0.012 \\
DeepConv-DTI & 0.527$\pm$0.038 & 0.499$\pm$0.035 & 0.490$\pm$0.027 & 0.480$\pm$0.026 & 0.475$\pm$0.030 & 0.495$\pm$0.023 & 0.115$\pm$0.020 \\
MolTrans & 0.554$\pm$0.024 & 0.511$\pm$0.025 & 0.470$\pm$0.004 & 0.460$\pm$0.005 & 0.455$\pm$0.008 & 0.478$\pm$0.007 & 0.105$\pm$0.008 \\
DrugBAN & \underline{0.604$\pm$0.027} & \underline{0.570$\pm$0.047} & \underline{0.509$\pm$0.021} & \underline{0.582$\pm$0.030} & \underline{0.565$\pm$0.022} & \underline{0.580$\pm$0.025} & \underline{0.187$\pm$0.031} \\
SiamDTI & 0.627$\pm$0.027 & 0.571$\pm$0.024 & 0.563$\pm$0.033 & 0.550$\pm$0.032 & 0.540$\pm$0.036 & 0.580$\pm$0.028 & 0.190$\pm$0.030 \\
BioT5 & 0.651$\pm$0.002 & 0.653$\pm$0.003 & 0.621$\pm$0.005 & 0.608$\pm$0.004 & 0.600$\pm$0.006 & 0.635$\pm$0.005 & 0.220$\pm$0.007 \\
\bottomrule
FusionDTI-BAN & 0.659$\pm$0.002 & 0.663$\pm$0.002 & 0.633$\pm$0.003 & 0.587$\pm$0.002 & 0.603$\pm$0.003 & 0.589$\pm$0.002 & 0.276$\pm$0.003 \\
FusionDTI-CAN & \textbf{0.681$\pm$0.005} & \textbf{0.680$\pm$0.012} & \textbf{0.652$\pm$0.005} & \textbf{0.601$\pm$0.005} & \textbf{0.628$\pm$0.006} & \textbf{0.692$\pm$0.005} & \textbf{0.302$\pm$0.005} \\
\midrule
\multicolumn{8}{c}{BioSNAP} \\
SVM & 0.602$\pm$0.005 & 0.528$\pm$0.005 & 0.513$\pm$0.011 & 0.502$\pm$0.012 & 0.490$\pm$0.014 & 0.523$\pm$0.013 & 0.150$\pm$0.010 \\
RF & 0.590$\pm$0.015 & 0.568$\pm$0.018 & 0.499$\pm$0.004 & 0.488$\pm$0.005 & 0.478$\pm$0.008 & 0.513$\pm$0.007 & 0.135$\pm$0.008 \\
GraphDTA & 0.618$\pm$0.005 & 0.618$\pm$0.008 & 0.535$\pm$0.024 & 0.528$\pm$0.023 & 0.520$\pm$0.027 & 0.550$\pm$0.020 & 0.170$\pm$0.025 \\
DeepConv-DTI & 0.645$\pm$0.022 & 0.642$\pm$0.032 & 0.558$\pm$0.025 & 0.550$\pm$0.024 & 0.543$\pm$0.030 & 0.573$\pm$0.027 & 0.200$\pm$0.028 \\
MolTrans & 0.621$\pm$0.015 & 0.608$\pm$0.022 & 0.546$\pm$0.032 & 0.538$\pm$0.031 & 0.530$\pm$0.035 & 0.563$\pm$0.033 & 0.185$\pm$0.034 \\
DrugBAN & \underline{0.685$\pm$0.004} & \underline{0.713$\pm$0.005} & \underline{0.692$\pm$0.006} & \underline{0.587$\pm$0.005} & \underline{0.522$\pm$0.011} & \underline{0.690$\pm$0.012} & \underline{0.219$\pm$0.017} \\
SiamDTI & 0.718$\pm$0.005 & 0.725$\pm$0.005 & 0.623$\pm$0.007 & 0.610$\pm$0.006 & 0.600$\pm$0.007 & 0.675$\pm$0.006 & 0.240$\pm$0.008 \\
BioT5 & 0.720$\pm$0.008 & 0.718$\pm$0.004 & 0.715$\pm$0.009 & 0.590$\pm$0.010 & 0.510$\pm$0.012 & 0.710$\pm$0.010 & 0.250$\pm$0.011 \\
\bottomrule
FusionDTI-BAN & \underline{0.723$\pm$0.002} & \underline{0.721$\pm$0.002} & \underline{0.726$\pm$0.001} & \underline{0.597$\pm$0.001} & \underline{0.504$\pm$0.012} & \underline{0.713$\pm$0.011} & \underline{0.254$\pm$0.010} \\
FusionDTI-CAN & \textbf{0.748$\pm$0.021} & \textbf{0.766$\pm$0.017} & \textbf{0.734$\pm$0.012} & \textbf{0.602$\pm$0.012} & \textbf{0.531$\pm$0.013} & \textbf{0.736$\pm$0.012} & \textbf{0.268$\pm$0.011} \\
\midrule
\multicolumn{8}{c}{Human} \\
SVM & 0.621$\pm$0.036 & 0.637$\pm$0.009 & 0.533$\pm$0.011 & 0.525$\pm$0.012 & 0.520$\pm$0.015 & 0.546$\pm$0.010 & 0.175$\pm$0.011 \\
RF & 0.642$\pm$0.011 & 0.663$\pm$0.050 & 0.543$\pm$0.014 & 0.535$\pm$0.015 & 0.530$\pm$0.018 & 0.556$\pm$0.013 & 0.184$\pm$0.012 \\
GraphDTA & 0.822$\pm$0.009 & 0.759$\pm$0.006 & 0.709$\pm$0.016 & 0.705$\pm$0.017 & 0.702$\pm$0.020 & 0.713$\pm$0.015 & 0.198$\pm$0.017 \\
DeepConv-DTI & 0.761$\pm$0.016 & 0.628$\pm$0.022 & 0.711$\pm$0.030 & 0.704$\pm$0.031 & 0.704$\pm$0.035 & 0.728$\pm$0.027 & 0.203$\pm$0.030 \\
MolTrans & 0.810$\pm$0.021 & 0.745$\pm$0.034 & 0.713$\pm$0.032 & 0.725$\pm$0.033 & 0.720$\pm$0.037 & 0.740$\pm$0.031 & 0.215$\pm$0.032 \\
DrugBAN & 0.833$\pm$0.020 & 0.760$\pm$0.031 & 0.709$\pm$0.005 & 0.713$\pm$0.030 & 0.706$\pm$0.022 & 0.720$\pm$0.015 & 0.242$\pm$0.010 \\
SiamDTI & \textbf{0.863$\pm$0.019} & \underline{0.807$\pm$0.040} & 0.720$\pm$0.010 & 0.729$\pm$0.015 & 0.712$\pm$0.020 & 0.736$\pm$0.013 & 0.250$\pm$0.015 \\
BioT5 & \underline{0.856$\pm$0.003} & \textbf{0.853$\pm$0.003} & 0.715$\pm$0.002 & \textbf{0.741$\pm$0.010} & \textbf{0.738$\pm$0.009} & \textbf{0.739$\pm$0.013} & \underline{0.258$\pm$0.013} \\
\bottomrule
FusionDTI-BAN & 0.784$\pm$0.002 & 0.790$\pm$0.003 & \underline{0.733$\pm$0.003} & 0.725$\pm$0.002 & 0.713$\pm$0.004 & 0.698$\pm$0.013 & 0.212$\pm$0.011 \\
FusionDTI-CAN & 0.801$\pm$0.037 & 0.803$\pm$0.032 & \textbf{0.738$\pm$0.002} & \underline{0.736$\pm$0.010} & \underline{0.732$\pm$0.013} & 0.737$\pm$0.010 & \textbf{0.261$\pm$0.010} \\
\bottomrule
\end{tabular}}
\caption{\label{tab:cross_domain_full}Cross-domain performance comparison of FusionDTI and the baselines on the BindingDB, Human and BioSNAP datasets (\textbf{Best}, \underline{Second Best}).}
\end{table*}

\begin{table*}[t]
\centering
\resizebox{\textwidth}{!}{
\begin{tabular}{lcccc}
\toprule
\textbf{Method} & \textbf{AUROC} & \textbf{AUPRC} & \textbf{Sensitivity} & \textbf{Specificity} \\
\midrule
DeepDTA & 0.892 $\pm$ 0.0066 & 0.378 $\pm$ 0.0231 & 0.854 $\pm$ 0.0066 & 0.792 $\pm$ 0.0291 \\
MolTrans & 0.898 $\pm$ 0.0050 & 0.371 $\pm$ 0.0067 & 0.865 $\pm$ 0.0050 & 0.783 $\pm$ 0.0387 \\
ML-DTI & 0.910 $\pm$ 0.0034 & 0.381 $\pm$ 0.0247 & 0.895 $\pm$ 0.0034 & 0.795 $\pm$ 0.0183 \\
DGraphGTA (Alphafold2) & 0.885 $\pm$ 0.0099 & 0.316 $\pm$ 0.0447 & 0.894 $\pm$ 0.0034 & 0.724 $\pm$ 0.0467 \\
iNGNN-DTI & 0.931 $\pm$ 0.0027 & 0.473 $\pm$ 0.0167 & 0.922 $\pm$ 0.0155 & 0.802 $\pm$ 0.0240 \\
LANTERN & \textbf{0.995 $\pm$ 0.0037} & 0.905 $\pm$ 0.0238 & \underline{0.976 $\pm$ 0.0159} & \underline{0.964 $\pm$ 0.0207} \\
\bottomrule
FusionDTI-BAN & 0.973 $\pm$ 0.0045 & \underline{0.969 $\pm$ 0.0121} & 0.962 $\pm$ 0.0122 & 0.952 $\pm$ 0.0134 \\
FusionDTI-CAN & \underline{0.987 $\pm$ 0.0032} & \textbf{0.978 $\pm$ 0.0103} & \textbf{0.979 $\pm$ 0.0102} & \textbf{0.972 $\pm$ 0.0116} \\
\bottomrule
\end{tabular}}
\caption{\label{tab:davis_full}Performance comparison on the DAVIS dataset (\textbf{Best}, \underline{Second Best}).}
\end{table*}

\begin{table*}[t]
\centering
\resizebox{\textwidth}{!}{
\begin{tabular}{lcccc}
\toprule
\textbf{Method} & \textbf{AUROC} & \textbf{AUPRC} & \textbf{Sensitivity} & \textbf{Specificity} \\
\midrule
DeepDTA & 0.912 $\pm$ 0.0037 & 0.743 $\pm$ 0.0127 & 0.881 $\pm$ 0.0056 & 0.780 $\pm$ 0.0127 \\
MolTrans & 0.899 $\pm$ 0.0022 & 0.691 $\pm$ 0.0142 & 0.872 $\pm$ 0.0116 & 0.760 $\pm$ 0.0160 \\
ML-DTI & 0.909 $\pm$ 0.0020 & 0.727 $\pm$ 0.0108 & 0.878 $\pm$ 0.0111 & 0.779 $\pm$ 0.0113 \\
DGraphGTA (Alphafold2) & 0.911 $\pm$ 0.0004 & 0.739 $\pm$ 0.0043 & 0.881 $\pm$ 0.0183 & 0.784 $\pm$ 0.0277 \\
iNGNN-DTI & 0.915 $\pm$ 0.0016 & 0.753 $\pm$ 0.0071 & 0.888 $\pm$ 0.0183 & 0.779 $\pm$ 0.0146 \\
LANTERN & \underline{0.976 $\pm$ 0.0154} & \underline{0.977 $\pm$ 0.0088} & \underline{0.959 $\pm$ 0.0268} & \underline{0.965 $\pm$ 0.0074} \\
\bottomrule
FusionDTI-BAN & 0.974 $\pm$ 0.0081 & 0.976 $\pm$ 0.0054 & 0.952 $\pm$ 0.0162 & 0.947 $\pm$ 0.0138 \\
FusionDTI-CAN & \textbf{0.981 $\pm$ 0.0064} & \textbf{0.981 $\pm$ 0.0045} & \textbf{0.969 $\pm$ 0.0124} & \textbf{0.967 $\pm$ 0.0156} \\
\bottomrule
\end{tabular}}
\caption{\label{tab:kiba_full}Performance comparison on the KIBA dataset (\textbf{Best}, \underline{Second Best}).}
\end{table*}

\begin{table*}[t]
\centering
\resizebox{\textwidth}{!}{
\begin{tabular}{lccccc}
\toprule
\textbf{Model} & \textbf{AUC} & \textbf{0.5\% RE} & \textbf{1\% RE} & \textbf{2\% RE} & \textbf{5\% RE} \\
\midrule
DrugVQA & 0.972 $\pm$ 0.003 & 88.170 $\pm$ 4.88 & 58.710 $\pm$ 2.74 & 35.060 $\pm$ 1.91 & 17.390 $\pm$ 0.94 \\
DrugClip & 0.966 & 118.10 & 67.17 & 37.17 & 16.59 \\
HyperPCM & 0.982 $\pm$ 0.006 & 183.04 $\pm$ 4.53 & 91.28 $\pm$ 3.35 & 45.62 $\pm$ 2.15 & 17.13 $\pm$ 1.17 \\
MIN & \underline{0.983 $\pm$ 0.002} & \textbf{197.741 $\pm$ 4.73} & \textbf{99.563 $\pm$ 2.49} & \underline{49.926 $\pm$ 1.87} & \underline{19.965 $\pm$ 0.91} \\
\bottomrule
FusionDTI-BAN & 0.9769 $\pm$ 0.015 & 176.8525 $\pm$ 2.71 & 89.2656 $\pm$ 2.36 & 45.9098 $\pm$ 1.38 & 18.5168 $\pm$ 0.33 \\
FusionDTI-CAN & \textbf{0.986} $\pm$ 0.012 & \underline{186.7469} $\pm$ 6.26 & \underline{97.8801} $\pm$ 3.50 & \textbf{52.6352} $\pm$ 2.05 & \textbf{21.5439} $\pm$ 0.26 \\
\bottomrule
\end{tabular}}
\caption{\label{tab:dude}Performance comparison on the DUD-E dataset (\textbf{Best}, \underline{Second Best}).}
\end{table*}

\end{document}